\documentclass[a4paper,11pt]{article}
\pdfoutput=1

\usepackage{jheppub}

\usepackage{subfig}
\usepackage{xspace}
\usepackage[countmax]{subfloat}
\usepackage{slashed}
\usepackage{longtable}

\usepackage{booktabs}
\usepackage{comment}
\usepackage{mathtools}
\usepackage{bbm}
\usepackage{soul}

\newcommand{\sig}{$H\to b\bar b$}
\newcommand{\bkg}{$g\to b\bar b$}

\def\be{\begin{equation}}
\def\ee{\end{equation}}

\def\zcut{z_{\text{cut}}}

\DeclareRobustCommand{\Sec}[1]{Sec.~\ref{#1}}

\DeclareRobustCommand{\Fig}[1]{Fig.~\ref{#1}}

\DeclareRobustCommand{\Eq}[1]{Eq.~(\ref{#1})}

\DeclareRobustCommand{\Ref}[1]{Ref.~\cite{#1}}

\title{Novel Jet Observables from Machine Learning}

\author{Kaustuv Datta and Andrew J.~Larkoski}
\affiliation{Physics Department, Reed College, Portland, OR 97202, USA}

\emailAdd{dattak@alumni.reed.edu}
\emailAdd{larkoski@reed.edu}

\abstract{
Previous studies have demonstrated the utility and applicability of machine learning techniques to jet physics. In this paper, we construct new observables for the discrimination of jets from different originating particles exclusively from information identified by the machine. The approach we propose is to first organize information in the jet by resolved phase space and determine the effective $N$-body phase space at which discrimination power saturates. This then allows for the construction of a discrimination observable from the $N$-body phase space coordinates.  A general form of this observable can be expressed with numerous parameters that are chosen so that the observable maximizes the signal vs.~background likelihood. Here, we illustrate this technique applied to discrimination of $H\to b\bar b$ decays from massive $g\to b\bar b$ splittings.  We show that for a simple parametrization, we can construct an observable that has discrimination power comparable to, or better than, widely-used observables motivated from theory considerations.  For the case of jets on which modified mass-drop tagger grooming is applied, the observable that the machine learns is essentially the angle of the dominant gluon emission off of the $b\bar b$ pair.
}

\begin{document} 
\maketitle

\section{Introduction}\label{sec:intro}

Several groups have recently applied promising machine learning techniques to the problem of classifying jets from different originating particles \cite{Cogan:2014oua,Almeida:2015jua,deOliveira:2015xxd,Baldi:2016fql,Guest:2016iqz,Conway:2016caq,Barnard:2016qma,Komiske:2016rsd,deOliveira:2017pjk,Kasieczka:2017nvn,Louppe:2017ipp,Dery:2017fap,Pearkes:2017hku,Cohen:2017exh,Butter:2017cot,Metodiev:2017vrx,Chang:2017kvc}.  A review of the advances of the field is presented in Ref.~\cite{Larkoski:2017jix}. These approaches, while demonstrating exceptional discrimination power, often come with the associated costs of utilizing hundreds of low-level input variables with thousands of correlations between them, and lack an immediately accessible physical interpretation. Ref.~\cite{Datta:2017rhs} presented the first application of a bottom-up organizing principle, whereby neural networks were trained and tested on minimal and complete bases of observables sensitive to the phase space of $M$ subjets in a jet. By appropriately identifying $M$ subjets in a jet, this study probed their phase space with sets of $3M-4$ infrared and collinear (IRC) safe observables. This essentially utilizes the distribution of the $M$ subjets on the phase space to identify useful information for discrimination. By increasing the dimensionality of the phase space in a systematic way, Ref.~\cite{Datta:2017rhs} used machine learning to demonstrate that in the case of discriminating boosted hadronic decays of $Z$ bosons from jets initiated by light partons, once 4-body phase space is resolved, no more information is observed to contribute meaningfully to discrimination power. In addition, Ref.~\cite{Aguilar-Saavedra:2017rzt} also presented a first application of this method to developing promising generic anti-QCD taggers that match or outperform the discrimination power of dedicated taggers.  Such approaches motivate the development of novel observables that can capture all of the salient information for discrimination of jets as learned by the machine. 

While the output of a neural network, boosted decision tree, or other machine learning method is itself an observable, it is in general a highly non-linear function of the input.  Additionally, the precise form of the explicit observable constructed by the machine is very sensitive to the assumed parameters; for example, the number of nodes or layers in a neural network.  In this paper, we propose a procedure to identify discriminating features of jets learned by machines to then generate novel observables.  These observables capture the important physics identified by the machine while at the same time being human-parseable. The general procedure is as follows:
\begin{enumerate}

 \item Construct a basis of observables that is sensitive to the phase space of subjets in a jet.  Measure these basis observables on your signal and background samples.
 
 \item Use machine learning techniques, such as neural networks, to identify the resolved $M$-body phase space at which signal vs.~background discrimination power saturates.
 
 \item Construct a function of the phase space variables (with tunable parameters) at which discrimination power saturates.  This function will be a new observable on the jets that can be used individually for discrimination.
 
 \item Fix the parameters in the new observable by demanding that it maximizes some discrimination metric, such as the area under the signal vs.~background efficiency curve (ROC curve).

\end{enumerate}
This algorithm is simple enough that it can be automated with essentially no human input, with a specified basis of observables to span $M$-body phase space and an appropriate functional form for the observable.  We will present and use a particular choice for the phase space basis and functional form of the final observable in this paper, but these may need to be modified and optimized for different studies.

For concreteness, here we apply the above approach to the problem of discriminating highly boosted decays of the Standard Model Higgs boson to a pair of $b$-quarks from splittings of gluons to $b$-quarks.  We study identification of $H\to b\bar b$ decays here as the signal and background jets both have a two-prong substructure, and theoretically-optimized discrimination observables have not been studied in great detail.  Recently, Ref.~\cite{CMS:2017cbv} utilized jet substructure approaches to propose a promising search strategy for this boosted decay mode of the Higgs, encouraging the possibility of discovery in data from Run-II of the LHC. In order to further increase the probability of discovery, it is necessary to explore new strategies to ensure sensitivity to the specific features of this decay mode. 

Using the organizing principle proposed in Ref.~\cite{Datta:2017rhs}, if discrimination power saturates at $M$-body phase space then the machine must be learning some function of the corresponding $3M-4$ phase space variables.  To resolve the $M$-body phase space, we use the $N$-subjettiness observables \cite{Thaler:2010tr,Thaler:2011gf}, as employed in Ref.~\cite{Datta:2017rhs}.  In the case of discrimination of boosted $H\to b\bar b$ decays to $g\to b\bar b$ splittings, we find that the discrimination power increases only slightly once 3-body phase space is resolved.  Thus, we will only study the resolved 3-body phase space in this paper.  The function of the observable on 3-body phase space that we study is a simple product form
\begin{equation}
\beta_{3}\equiv{\left(\tau_{1}^{(0.5)}\right)}^{a} {\left(\tau_{1}^{(1)}\right)}^{b} {\left(\tau_{1}^{(2)}\right)}^{c} {\left(\tau_{2}^{(1)}\right)}^{d} {\left(\tau_{2}^{(2)}\right)}^{e}.
\end{equation}
Here, the $\tau_{N}^{(\beta)}$ are the $N$-subjettiness observables, 3-body phase space is 5 dimensional, and the parameters $a,b,c,d,e$ will be chosen to maximize discrimination power.  We emphasize that while this product form is simple, there may be a better choice for the form of function on phase space.

\begin{figure}[]
	\begin{center}
		
		\subfloat[]{\label{fig:SD_nbody}
			\includegraphics[width=7.2cm]{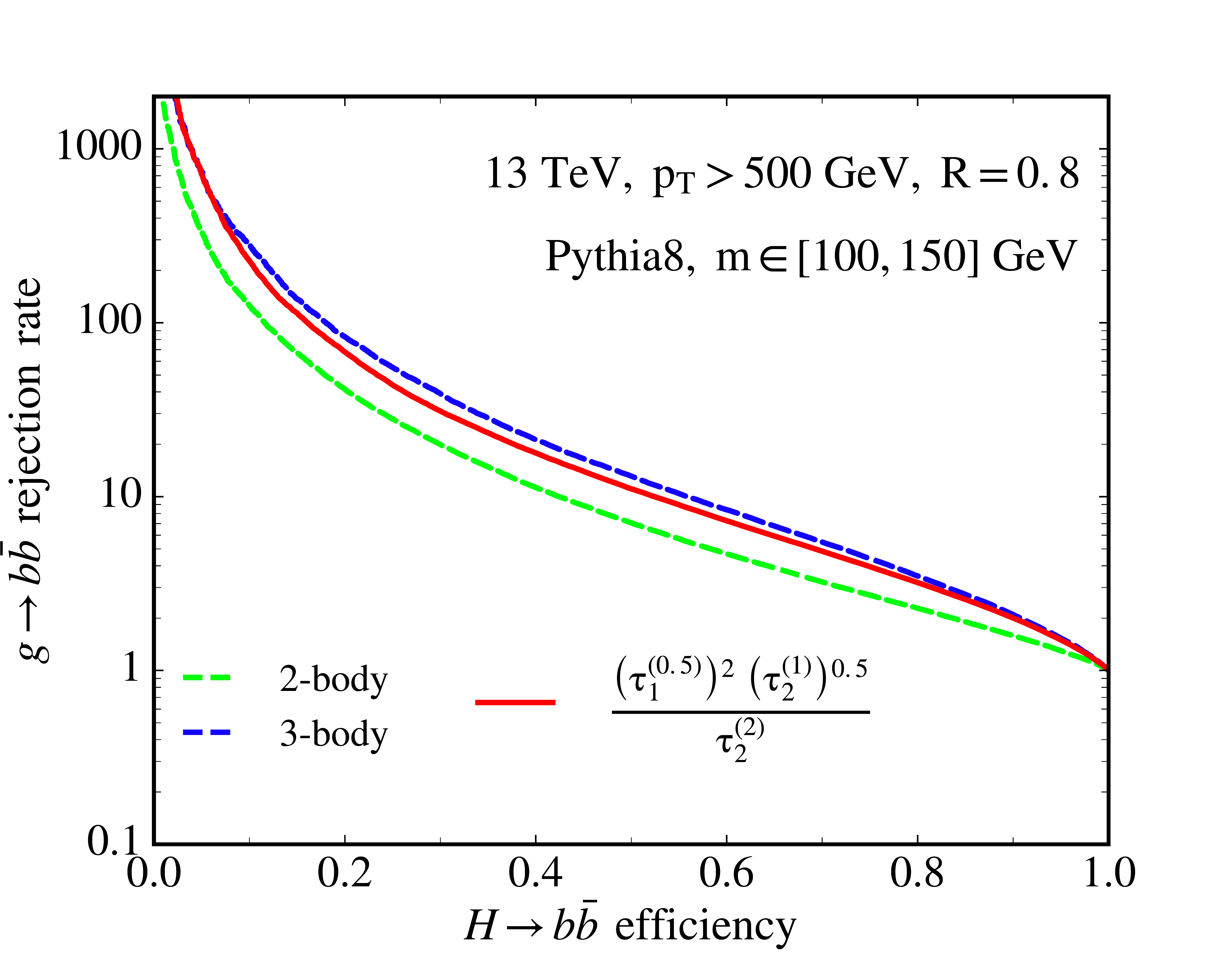}    
		}\qquad
		\subfloat[]{\label{fig:SD_NS}
			\includegraphics[width=7.2cm]{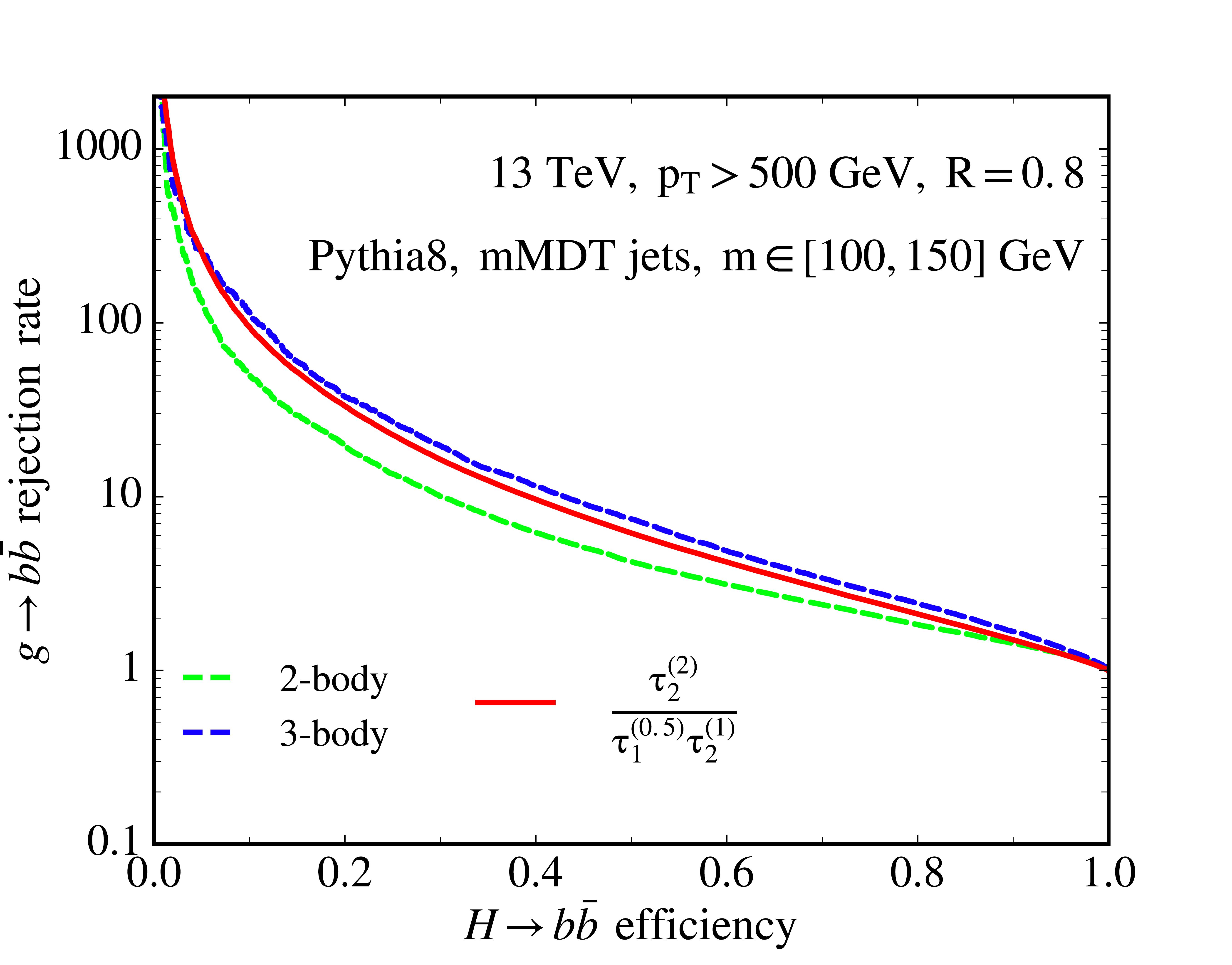}
		}
	
	\caption{
		\sig~ jet efficiency vs.~\bkg~ jet rejection rate plots for ungroomed (a) and groomed (b) jets.  For both, the $M$-body curves determined by a neural network demonstrate a large increase in discrimination power between 2- and 3-body phase space. Also shown (in red) is the new observable which captures the majority of information important for discrimination identified by resolving 3-body phase space.		
	}\label{fig:moneyplot}
\end{center}
\end{figure}		

We show in Fig.~\ref{fig:moneyplot} the results of this analysis.  We consider jets in simulation on which no grooming has been applied and on which the modified mass-drop tagger (mMDT) \cite{Dasgupta:2013ihk, Dasgupta:2013via} has been applied.  We then measure the mass $m_J$ of these ungroomed or groomed jets (as appropriate) and make a cut of $m_J\in [100,150]$ GeV, in the range of the Higgs peak.  On these jets, we then measure the 2- and 3-body phase space variables and determine the single function of the 3-body phase space variables as described above.  The signal and background efficiencies in Fig.~\ref{fig:moneyplot} do not include the effects of the mass cut, so that the curves end at 100\% signal and background efficiency.  The ROC curve for the new observable is shown as the solid line on these plots, and exhibits significant improvement over the 2-body phase space observables and nearly captures all discrimination power of 3-body phase space.  In the case of mMDT jets, we will show that this new observable effectively corresponds to the angle of the dominant gluon emission off of the $b\bar b$ pair.

The outline of this paper is as follows.  In \Sec{sec:basis} we review and define the minimal and complete observable bases that are used to identify the coordinates of $M$-body phase space, and discuss the mMDT groomer.  In \Sec{sec:imp}, we describe our event simulation and machine learning implementation, and demonstrate the application of our procedure for developing the new observable in the case of \sig~vs.~\bkg. Further, we explore the physics implications the functional form of the observable, and compare its discrimination power to standard observables motivated from QCD considerations.  We conclude in \Sec{sec:conc} and discuss other possible applications of this  procedure.  

\section{Observable basis }\label{sec:basis}

In this section, we discuss the basis of IRC safe observables that we use to identify structure in the jet, following the approach presented in Ref.~\cite{Datta:2017rhs}.  For our analysis, we exclusively use the $N$-subjettiness observables \cite{Stewart:2010tn,Thaler:2010tr,Thaler:2011gf}. This is without loss of generality and the analysis can, for example, be equivalently implemented with the $N$-point energy correlation functions \cite{Larkoski:2013eya} or the four-momentum of subjets from the exclusive $k_T$ algorithm.  This specific choice for each $M$-body basis is only to ensure that the set of observables minimally and completely span the phase space of emissions in a jet.   

The $N$-subjettiness observable $\tau_N^{(\beta)}$ provides a measure of the radiation about $N$ axes in the jet, specified by the angular exponent $\beta>0$:
\begin{equation}
\tau_N^{(\beta)} = \frac{1}{p_{T J}} \sum_{i\in \text{Jet}} p_{Ti} \min\left\{
R_{1i}^\beta,R_{2i}^\beta,\dotsc,R_{Ni}^\beta
\right\}\,.
\end{equation}
Here, $p_{TJ}$ is the transverse momentum of the jet of interest, $p_{Ti}$ is the transverse momentum of particle $i$ in the jet, and $R_{Ki}$, for $K=1,2,\dotsc,N$, is the angle in pseudorapidity and azimuth between particle $i$ and axis $K$ in the jet.  In our analyses, we choose to define the $N$ axes in the jet according to the exclusive $k_T$ algorithm \cite{Catani:1993hr,Ellis:1993tq} with $E$-scheme recombination \cite{Blazey:2000qt}.  

To identify structure in the jet, we use the organizing principle proposed in Ref~\cite{Datta:2017rhs} so that our choice of basis of observables is complete and minimal.  We first identify the set of $N$-subjettiness observables that completely specify the coordinates of $M$-body phase space.  Since $M$-body phase space is $3M-4$ dimensional, we only measure $3M-4$ $N$-subjettiness observables, as follows:
\begin{equation}
\left\{
\tau_1^{(0.5)},\tau_1^{(1)},\tau_1^{(2)},\tau_2^{(0.5)},\tau_2^{(1)},\tau_2^{(2)},\dotsc,\tau_{M-2}^{(0.5)},\tau_{M-2}^{(1)},\tau_{M-2}^{(2)},\tau_{M-1}^{(1)},\tau_{M-1}^{(2)}
\right\}\,.
\end{equation}
For further details on this method, we ask the reader to refer to Ref.~\cite{Datta:2017rhs}. Note that when all particles have non-zero energy and are at a finite angle to one another, the $3(M-2)+2=3M-4$ observables span the space of phase space variables for generic momenta configurations.  Following from the above, we list the sets of observables that were used in our analysis for resolving particular $M$-body phase space:
\begin{align*}
	&\text{2-body: } \tau_1^{(1)}\,,\tau_1^{(2)}\\
	&\text{3-body: } \tau_1^{(0.5)}\,,\tau_1^{(1)}\,,\tau_1^{(2)}\,,\tau_2^{(1)}\,,\tau_2^{(2)}\\
	&\text{4-body: } \tau_1^{(0.5)}\,,\tau_1^{(1)}\,,\tau_1^{(2)}\,,\tau_2^{(0.5)}\,,\tau_2^{(1)}\,,\tau_2^{(2)}\,,\tau_3^{(1)}\,,\tau_3^{(2)}\\
	&\text{5-body: } \tau_1^{(0.5)}\,,\tau_1^{(1)}\,,\tau_1^{(2)}\,,\tau_2^{(0.5)}\,,\tau_2^{(1)}\,,\tau_2^{(2)}\,,\tau_3^{(0.5)}\,,\tau_3^{(1)}\,,\tau_3^{(2)}\,,\tau_4^{(1)}\,,\tau_4^{(2)}\\
	&\text{6-body: } \tau_1^{(0.5)}\,,\tau_1^{(1)}\,,\tau_1^{(2)}\,,\tau_2^{(0.5)}\,,\tau_2^{(1)}\,,\tau_2^{(2)}\,,\tau_3^{(0.5)}\,,\tau_3^{(1)}\,,\tau_3^{(2)}\,,\tau_4^{(0.5)}\,,\tau_4^{(1)}\,,\tau_4^{(2)}\,,\tau_5^{(1)}\,,\tau_5^{(2)}
\end{align*}

\subsection{mMDT Grooming Algorithm}
In the analysis of the next section, we measure the aforementioned observables on samples of both ungroomed jets and jets groomed with the modified mass-drop tagger (mMDT) \cite{Dasgupta:2013ihk,Dasgupta:2013via}. Given a set of constituents of a jet with radius $R$, and for a fixed transverse momentum fraction parameter $\zcut$, the mMDT grooming algorithm proceeds as follows:
\begin{enumerate}
	\item Recluster the jet with the Cambridge/Aachen (C/A) algorithm \cite{Dokshitzer:1997in,Wobisch:1998wt,Adloff:2000tq}.
	
	\item Sequentially step through the branching history of the reclustered jet. At each branching with daughter branches $i$ and $j$, check the mMDT criterion 
	\begin{equation}
	\frac{\min(p_{Ti},p_{Tj})}{p_{Ti}+p_{Tj}}>\zcut.
	\end{equation}
	If the condition fails, drop the softer of two daughter branches and follow through to the next branching in the rest of the clustering history. 
	\item This continues until the mMDT criterion is passed. At this point the algorithm terminates and the final jet is groomed of all branches that fail to pass the test. This ensures that the softest emissions at wide angles from the hard subjets, and contamination from the underlying event (UE) and initial state radiation (ISR), are effectively removed from the final groomed jet.
\end{enumerate}
In the groomed case, the observable bases are measured after grooming, thus the collection of particles in the jet effectively contributing to the phase space are different than for ungroomed jets.

\section{How to Make an Observable}\label{sec:imp}

In this section, we describe our event simulation and implementation of machine learning to the $N$-subjettiness basis of observables described in the previous section.  We generate background $pp\to Z+b\bar{b}$  and signal $pp\to Z(H\to b\bar b)$ events at the 13 TeV LHC with MadGraph5 v2.5.4 \cite{Alwall:2014hca}.  The $Z$ bosons were used as a control and decayed exclusively to neutrinos.  These tree-level events are then showered in Pythia v8.226 \cite{Sjostrand:2006za,Sjostrand:2014zea} with default settings.  Later in this section we will also show results obtained by applying the observable learned from Pythia to events showered with Herwig v7.1.1 \cite{Bahr:2008pv,Bellm:2015jjp}. We use FastJet v3.2.1 \cite{Cacciari:2011ma,Cacciari:2005hq} to cluster the jets.  On the clustered anti-$k_T$ \cite{Cacciari:2008gp} jets with radius $R=0.8$ and minimum $p_T$ of 500 GeV, we then measure the basis of $N$-subjettiness observables using the code provided in FastJet contrib v1.026.   The observables are measured at the particle level and we do not apply any detector simulation. 

We then study these jets without grooming and with mMDT grooming with $\zcut = 0.1$.  On these samples, we measure the jet mass and apply a cut of $100 < m_J < 150$ GeV which selects the Higgs signal region. Additionally, we measure the sufficient collection of $N$-subjettiness observables to completely determine up through 6-body phase space.  We then proceed to develop novel observables learned from the machine that further discriminate signal and background.

\begin{figure}[]
	\begin{center}
		\subfloat[]{\label{fig:SD_DNN}
			\includegraphics[width=7.2cm]{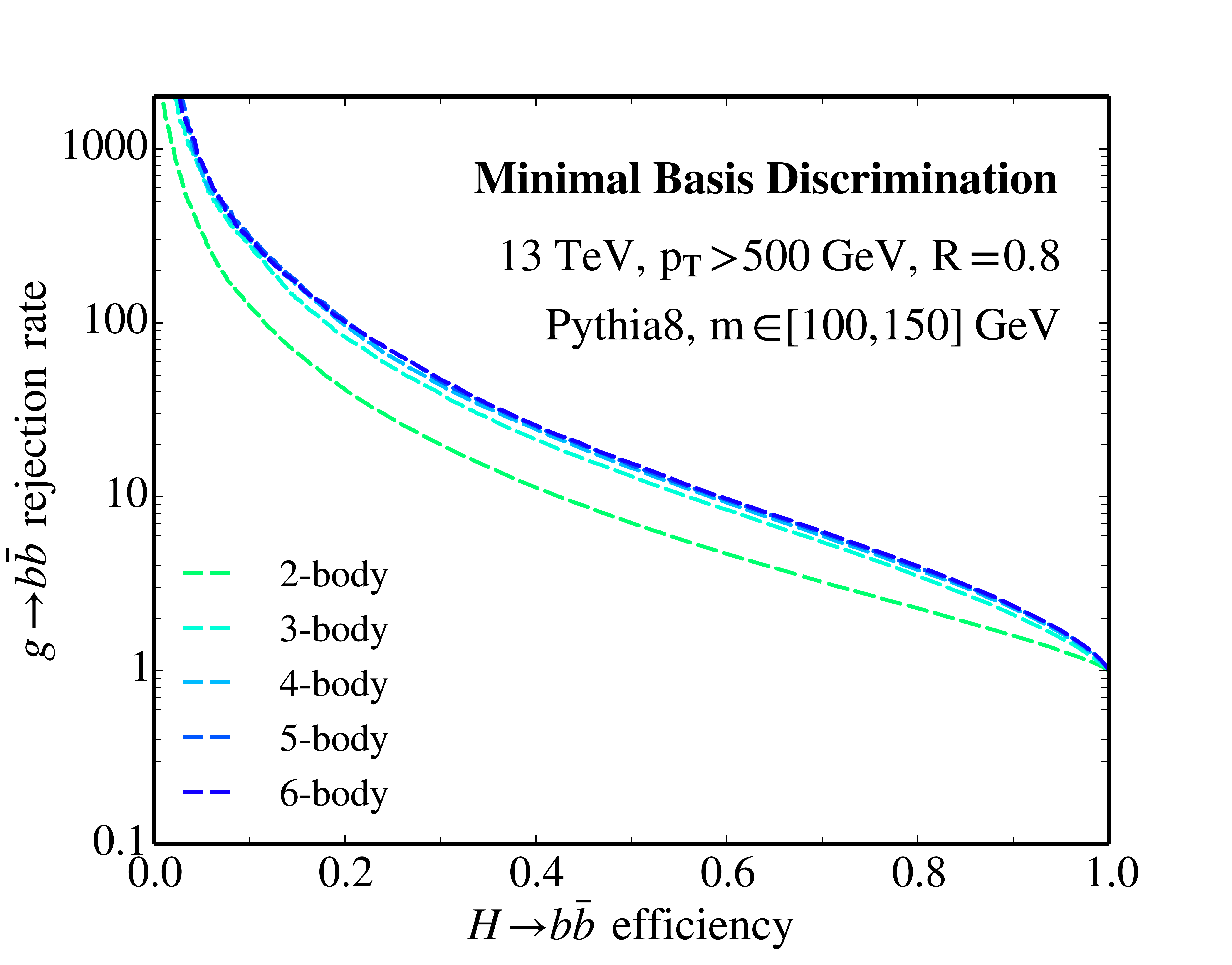}    
		}\qquad
		\subfloat[]{\label{fig:nSD_DNN}
			\includegraphics[width=7.2cm]{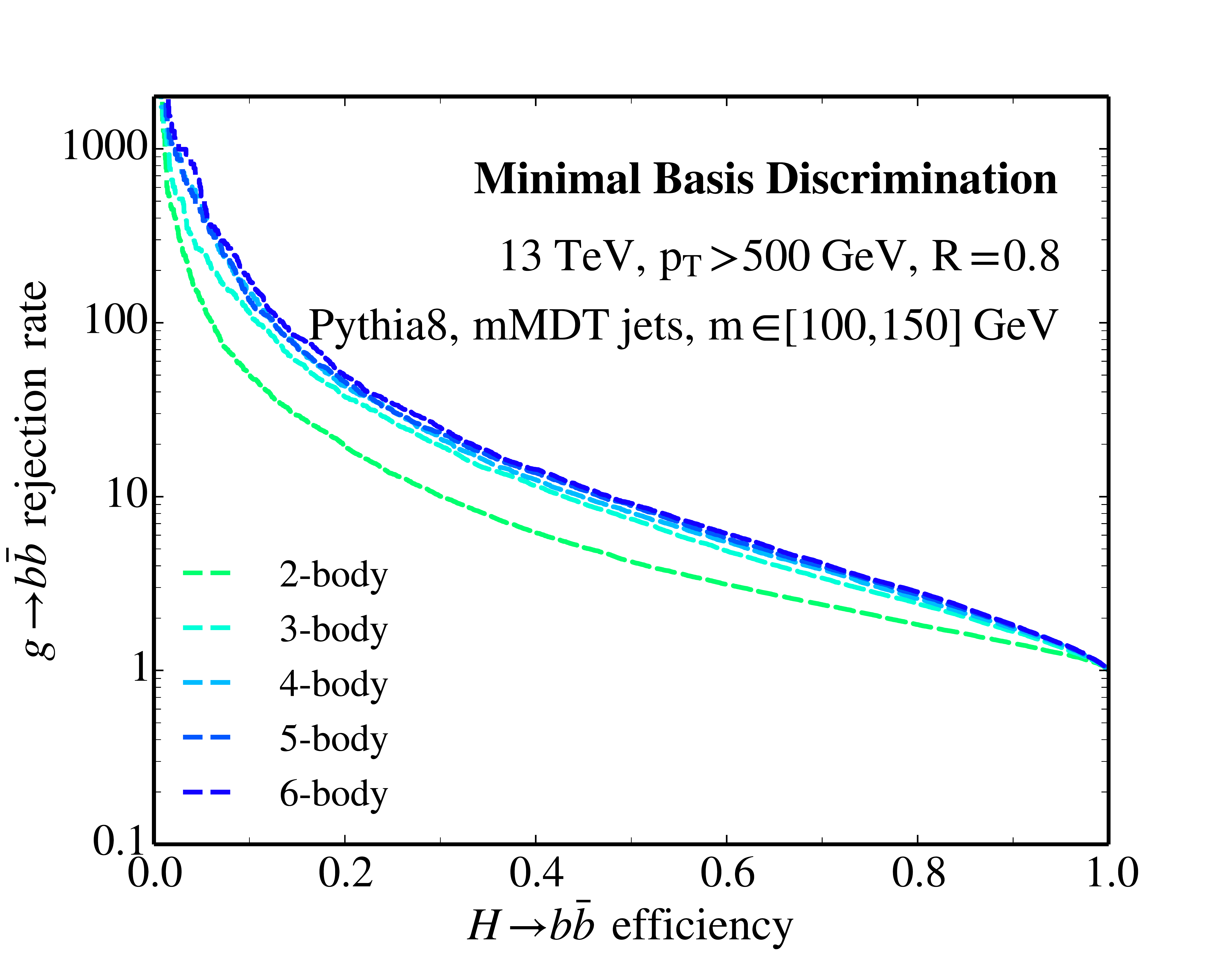}
		}\\
	\end{center}
	\caption{
		\sig~ jet efficiency vs. \bkg~ jet rejection rate plot for the ungroomed (a) and groomed (b) jets, as determined by a neural network. The curves effectively demonstrate saturation of discrimination power on the resolution of 3-body phase space.}
	\label{fig:satplot}
\end{figure}	

To do this, we use the approach of Ref.~\cite{Datta:2017rhs}.  This enables us to identify the resolved phase space that captures the vast majority of the discrimination power.  To calculate the ROC curves for $M$-body phase space shown in Fig.~\ref{fig:satplot}, we trained deep neural networks with fully-connected layers on the bases of $N$-subjettiness observables discussed in the previous section.  Discrimination power is seen to dramatically increase on going from 2- to 3-body phase space, and higher phase space improves discrimination only slightly. All networks were trained using the Keras \cite{chollet2015keras} deep learning libraries.  However, it is important to note here that this procedure, as justified in Ref.~\cite{Datta:2017rhs}, is agnostic of the specific machine learning method used and, equivalently, other machine learning methods (like a boosted decision tree) could have been used to identify the point of saturation.  From these results, in what follows we will exclusively study 3-body phase space.

From the assumption that 3-body phase space effectively saturates discrimination power, our goal is to define a single observable that captures this discrimination power.  As previously discussed, this proposed observable must be a function of the phase space variables at the point of saturation.  Determining this function thus requires parametrizing the possible functions of the phase space variables somehow.  Our approach will be illustrative and demonstrate the procedure for doing so. However, there may be smarter or more effective ways to optimize this process.  Here, we will just consider the observable formed from the product of 3-body (5 dimensional) phase space variables, raised to powers $a,~b,~c,~d$ and $e$:
\begin{equation}
\beta_{3}={\left(\tau_{1}^{(0.5)}\right)}^{a} {\left(\tau_{1}^{(1)}\right)}^{b} {\left(\tau_{1}^{(2)}\right)}^{c} {\left(\tau_{2}^{(1)}\right)}^{d} {\left(\tau_{2}^{(2)}\right)}^{e}.
\end{equation}
At this stage, the optimal values of these powers are undetermined, and there is no guarantee that this form of the observable actually includes all discrimination power of 3-body phase space.  We leave the problem of a complete observable basis to future work.

\subsection{Determining Optimal Parameters}\label{subsec:obsgen}

Utilizing the measurements of $N$-subjettiness observables from our datasets of groomed and ungroomed jets, we run a Monte Carlo simulation whereby uniform random numbers in the range $[-5,5]$ were assigned to the exponents $a,~b,~c,~d$ and $e$.\footnote{We also attempted to identify the set of exponent values that maximizes the AUC using stochastic gradient descent.  However, due to the finite binning necessary to calculate the likelihood and therefore the AUC, we were unable to demonstrate satisfactory convergence to maxima.  Additionally, the Monte Carlo approach enables a direct study of correlations of the exponents on the discrimination power.  This will be demonstrated shortly.} In each run, values of the resultant product observable were measured on samples of 200,000 signal and background jets from Pythia that passed the mass cut of $m_J\in[100,150]$ GeV.  We then construct the 1 dimensional binned likelihood distributions of the observable, which is the optimal discriminant for a given functional form of the observable by the Neyman-Pearson lemma \cite{Neyman289}.  The likelihood distributions of the observable, for each set of exponent values, were then used to calculate the area under the ROC curve (AUC) to estimate the discrimination power.  For each run, the values of the exponents and the calculated AUC were stored only when the  AUC crossed a threshold value of 0.5. This was necessary to exclude binning effects on the measured discrimination power of the observable.

\begin{figure}[]
	\begin{center}
		\subfloat[]{\label{fig:hist_SD}
			\includegraphics[width=7.2cm]{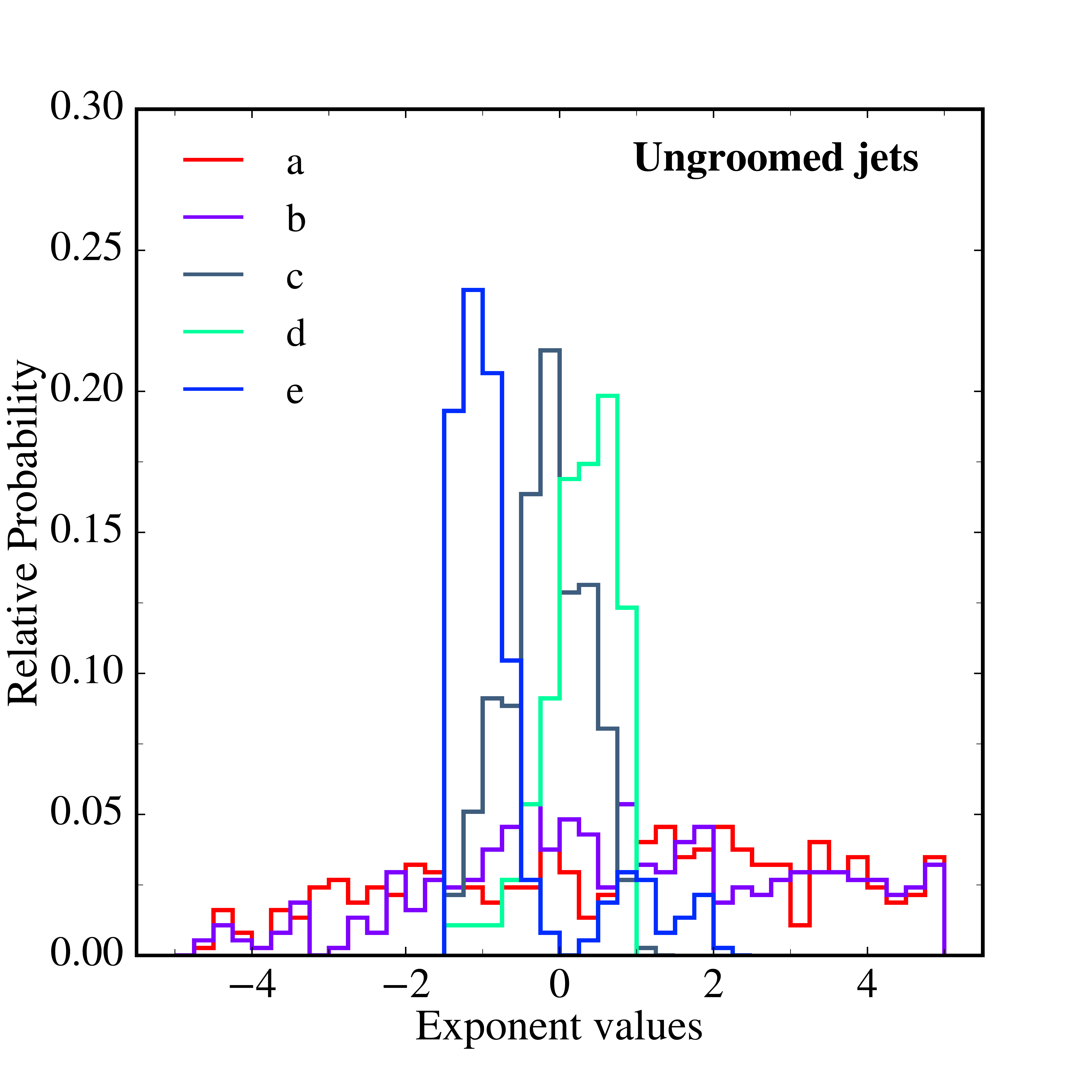}    
		}\qquad
		\subfloat[]{\label{fig:hist}
			
			\includegraphics[width=7.2cm]{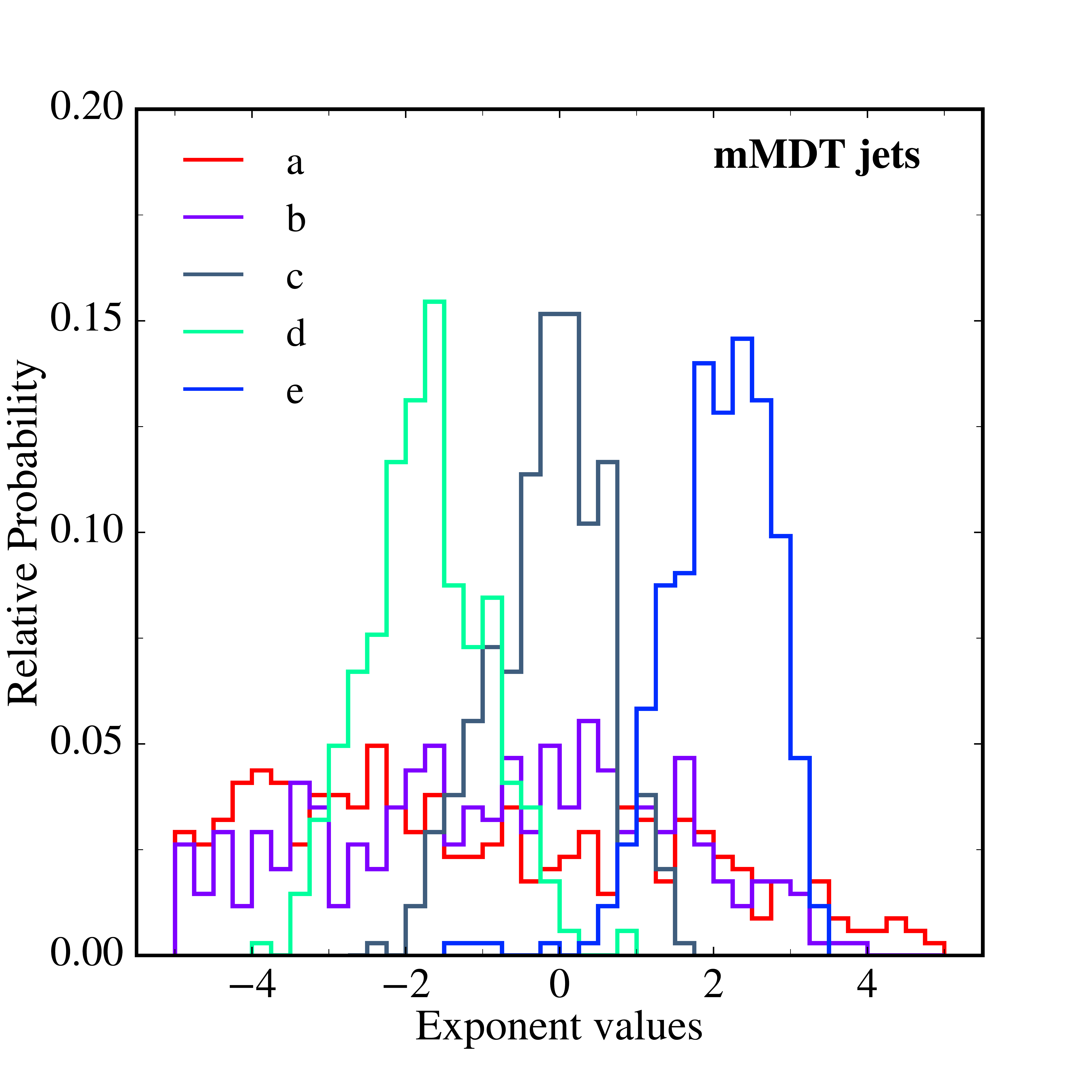}
		}
	\end{center}
	\caption{
		Histograms for the values of exponents of the product observable. For ungroomed (a) and groomed (b) jets exponent values in these histograms were accepted when the generated AUCs for the binned signal and background likelihood distributions were above 0.81 and 0.73 respectively.  
	}
	\label{fig:exp_hist}
\end{figure}

We apply this procedure to jets that have been groomed with mMDT and those that have not.  In the groomed case, due to the exclusion of soft emissions and contamination from initial state radiation (ISR) or underlying event, it is relatively straightforward to extract a useful physical understanding from the obtained functional form of the observable.  In Fig.~\ref{fig:exp_hist}, we plot the distributions of the exponents $a,~b,~c,~d$ and $e$ with the requirement that the AUC for the corresponding product observable is greater than 0.81 or 0.73, for ungroomed and groomed jets, respectively.  These distributions will enable us to extract the exponent values for which the AUC is maximized for the binned likelihood distributions of the product observable measured on signal and background.

By studying the histograms of the exponents one can make the following conclusions:
	\begin{itemize}
		
		\item For the ungroomed jets, AUC is maximized when $c=0,~d=0.5,~e=-1$ as the distributions for these exponents are very narrow. Since the distributions for $a$ and $b$ are both approximately uniform on $[-5,5]$, further interpretation is required.  This will be done shortly.
		
		\item For the groomed jets, AUC is maximized when $c=0,~d=-2,~e=2$. Again, since $a$ and $b$ are both approximately uniformly distributed over $[-5,5]$, further analysis is required to determine the values that maximize the AUC.
	\end{itemize}
To determine the values for $a$ and $b$ for both ungroomed and groomed jets, we work to understand the correlation between the exponents. 

\begin{figure}[h]
	\begin{center}
		\subfloat[]{\label{fig:avb_sd}
			\includegraphics[width=7.2cm]{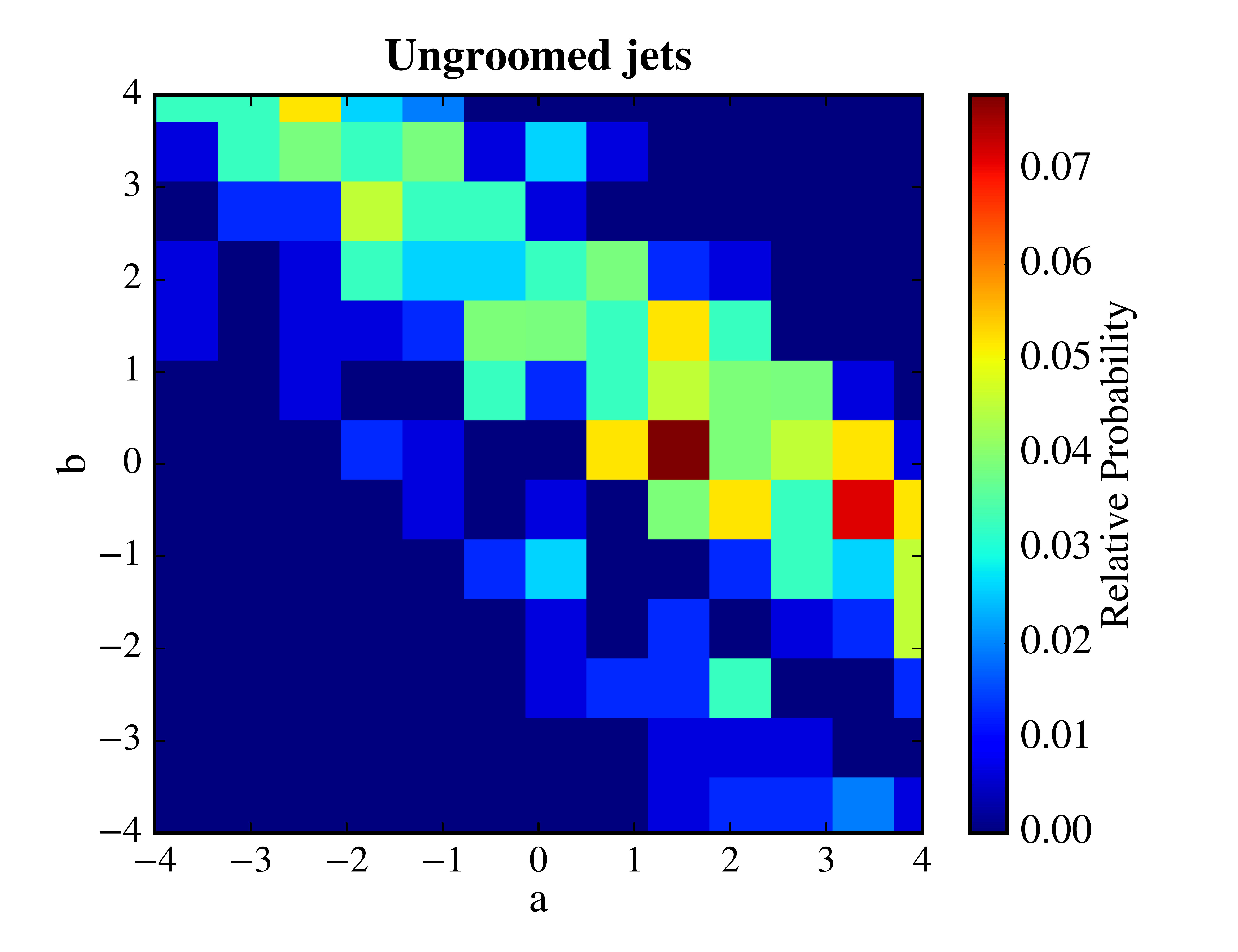}    
		}\qquad
		\subfloat[]{\label{fig:avb_nsd}
			\includegraphics[width=7.2cm]{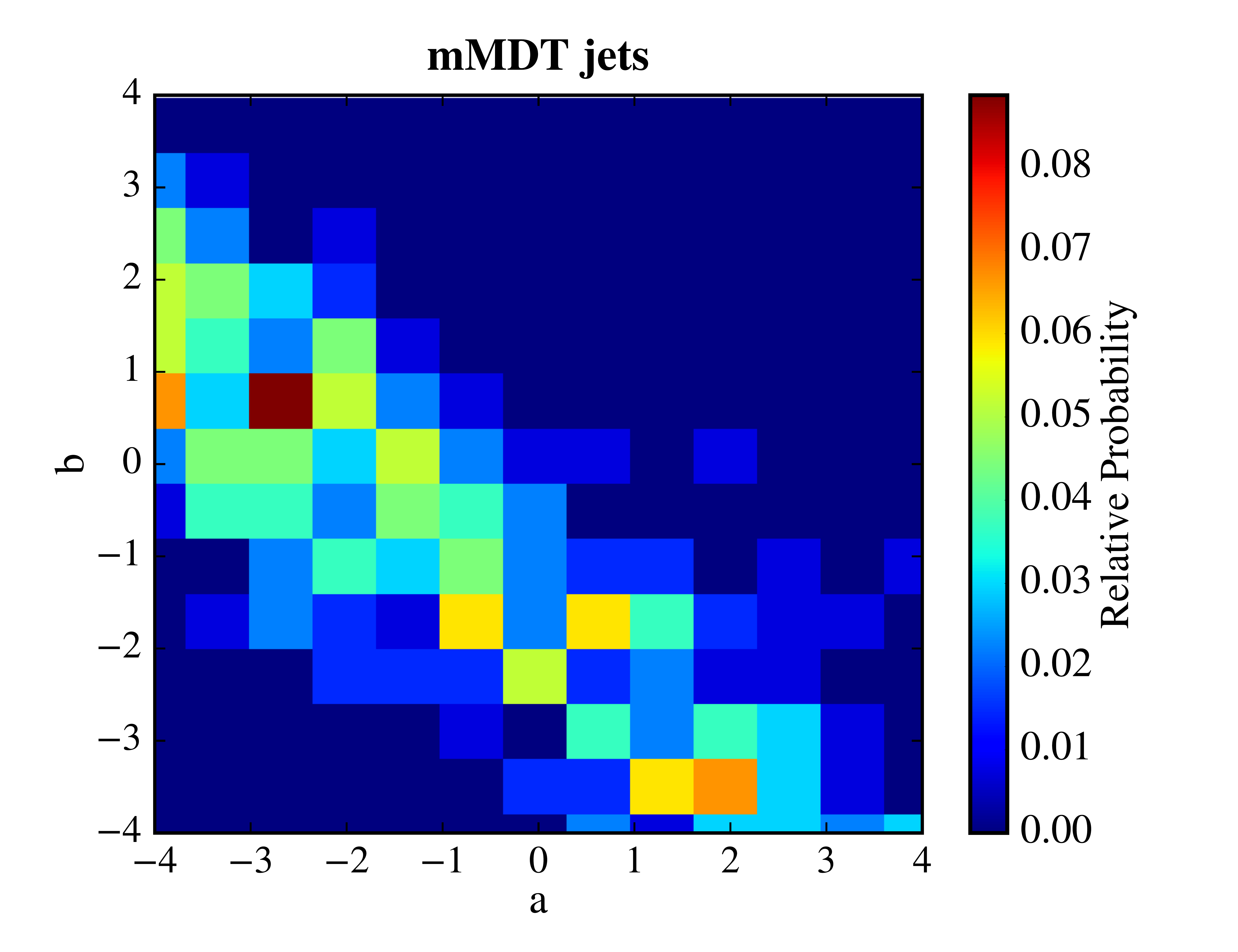}
		}
	\end{center}
	\caption{
		Heat maps of the correlation between $a$ and $b$ exponents of the product observable for ungroomed (a) and groomed (b) jets.
	}
	\label{fig:hist2d}
\end{figure}	

\begin{figure}[]
	\begin{center}
		
		\includegraphics[width=0.5\textwidth]{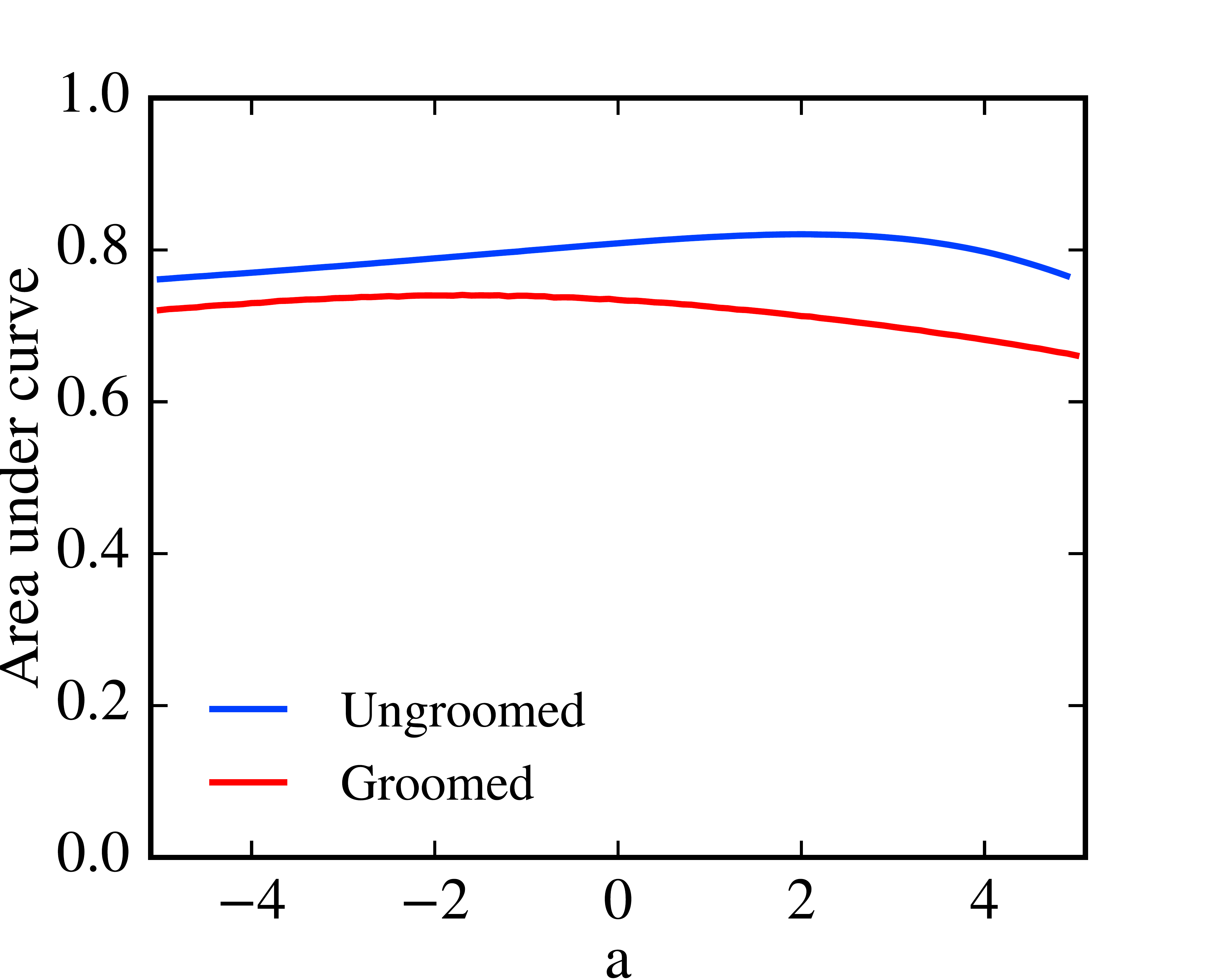}    
		
	\end{center}
	\caption{	
		Variation of area under the ROC curve for the observable when the $a$ exponent is scanned over the range $[-5,5]$, keeping the $c,~d$ and $e$ exponents fixed and varying $b$ with $a$ as per \Eq{eq:avb}. 
	}
	\label{fig:auc_vs_a}
\end{figure}

To determine the correlation between the exponents $a$ and $b$, we plot their joint probability distribution from the uniform sampling on $[-5,5]$ with the same cuts on the resulting observables' AUC.  For both ungroomed and groomed jets, this is shown in \Fig{fig:hist2d}.  These plots demonstrate a strong correlation between these exponents, which to very good approximation is:
\begin{align}
	&\mathrm{Ungroomed:~}a+b=2\,, &\mathrm{Groomed:~}a+b=-2\,. \label{eq:avb}
\end{align}
These relationships can be used to fix $b$, for example, as a function of exponent $a$.  To determine the value of the exponent $a$, we then fix $b,~c,~d,$ and $e$ as earlier, and calculate the AUC for $a\in[-5,5]$.  The results of this scan are shown in \Fig{fig:auc_vs_a} for ungroomed and groomed jets.  In particular, we note from the plot that AUC is maximized for the ungroomed case when $a=2$ and for the groomed case when $a=-2$.  This implies that the exponent $b=0$ for both cases, using \Eq{eq:avb}.  

Thus, the product observable takes on the following forms for the two kinds of jets: 
\begin{align}
	&\mathrm{Ungroomed:~}\beta_{3}=\frac{\left({\tau_{1}^{(0.5)}}\right)^{2}\left({\tau_{2}^{(1)}}\right)^{0.5}} { \tau_{2}^{(2)}}\,,&\mathrm{Groomed:~}\beta_{3}^{(g)}=\left(\frac{\tau_{2}^{(2)}}{\tau_{1}^{(0.5)} {\tau_{2}^{(1)}}}\right)^{2}\,. \label{eq:init_form}
\end{align}
Any monotonic function of the observable will produce the same discrimination power, and so we can simplify the expression for the groomed product observable.  For the product observable for groomed jets, we use the expression:
\begin{equation}\label{eq:prodobs}
\beta_{3}^{(g)}=\frac{\tau_{2}^{(2)}}{\tau_{1}^{(0.5)} {\tau_{2}^{(1)}}}.
\end{equation}
It is interesting to note that the observables from this method are Sudakov safe \cite{Larkoski:2013paa,Larkoski:2015lea} because they are formed from ratios of IRC safe observables.

\subsection{Physical Interpretation}\label{subsec:obsphys}
\begin{figure}[]
	\begin{center}
		\subfloat[]{\label{fig:obshist_SD}
			\includegraphics[width=7.2cm]{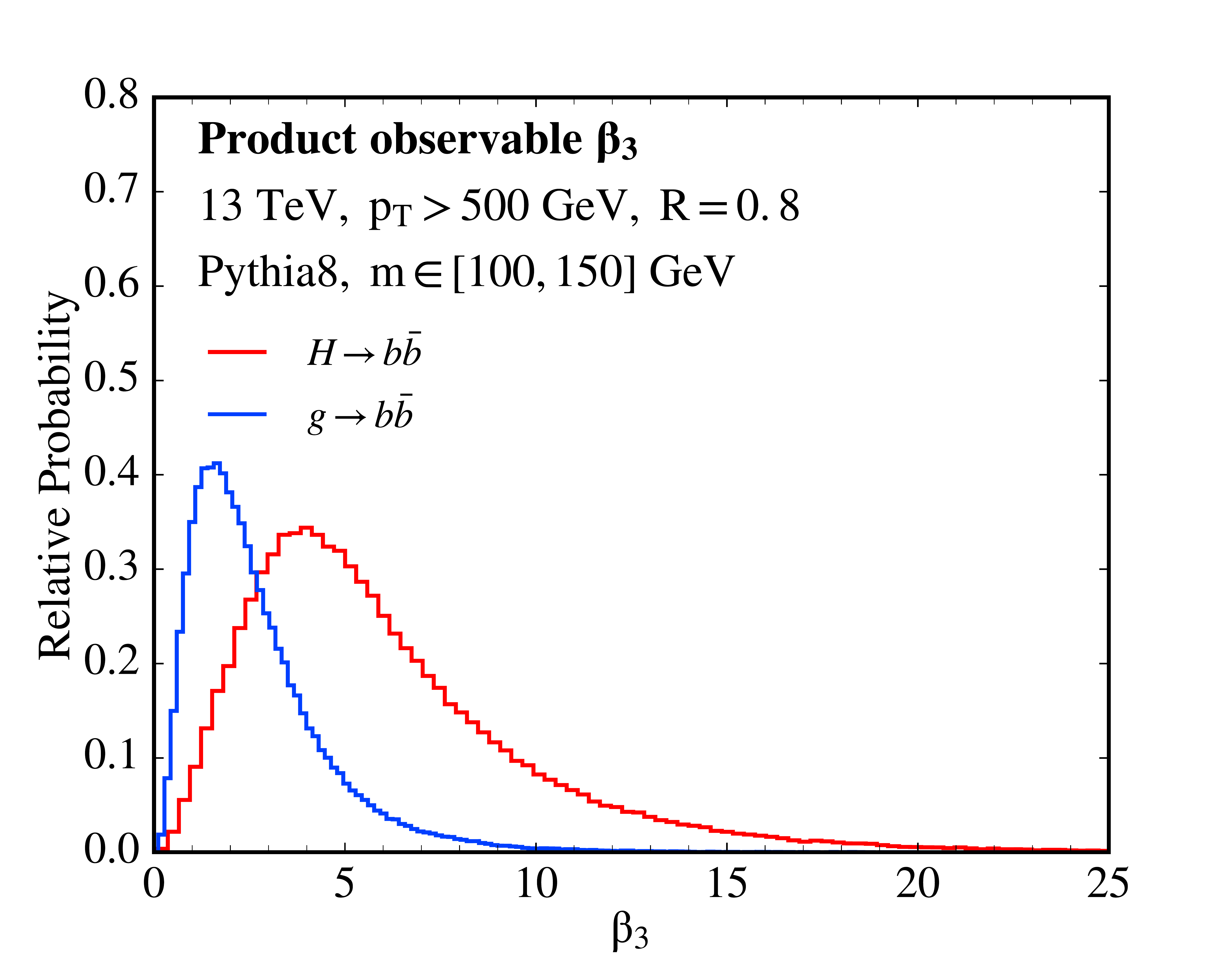}    
		}\qquad
		\subfloat[]{\label{fig:obshist}
			\includegraphics[width=7.2cm]{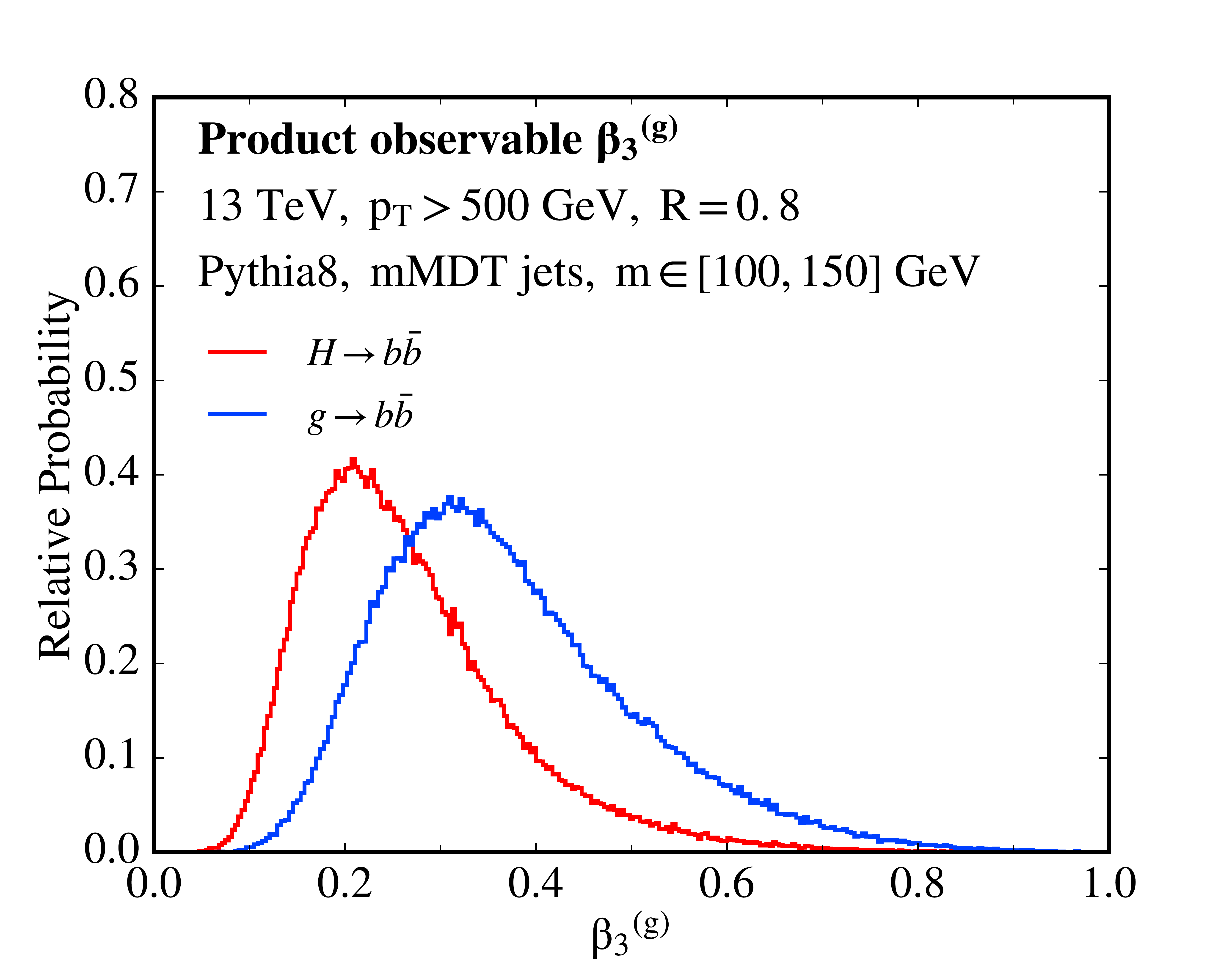}
		}\\
	\end{center}
	\caption{
		Distributions of the product observable for signal (red) and background (blue), measured on the samples of ungroomed (a) and groomed (b) jets showered with Pythia, within a mass cut of $m_J\in[100,150]$ GeV.  
	}
	\label{fig:prodobs}
\end{figure}	

\begin{figure}[]
	\begin{center}
		\subfloat[]{\label{fig:herwig_obshist_nSD}
			\includegraphics[width=7.2cm]{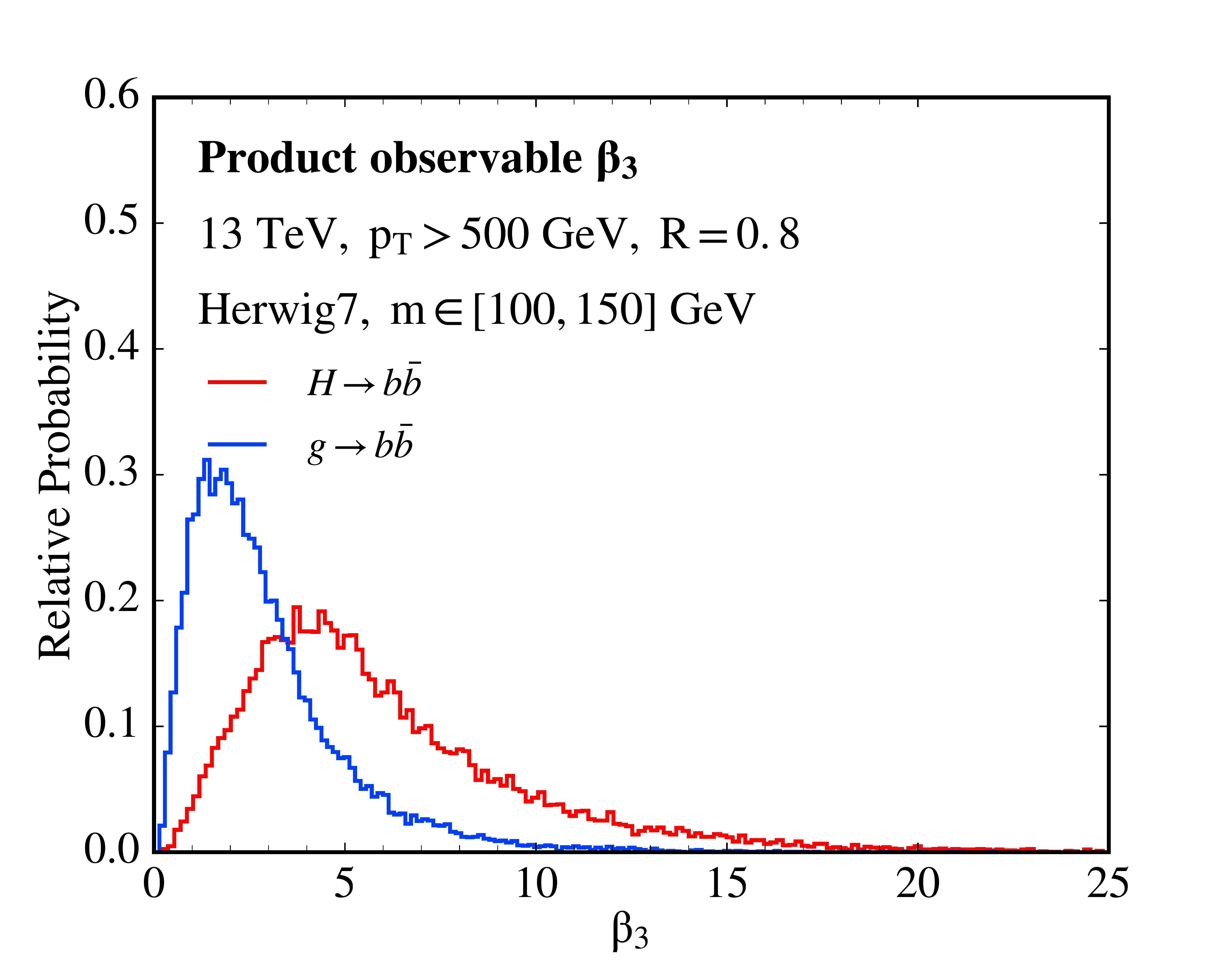}    
		}\qquad
		\subfloat[]{\label{fig:herwig_obshist_SD}
			\includegraphics[width=7.2cm]{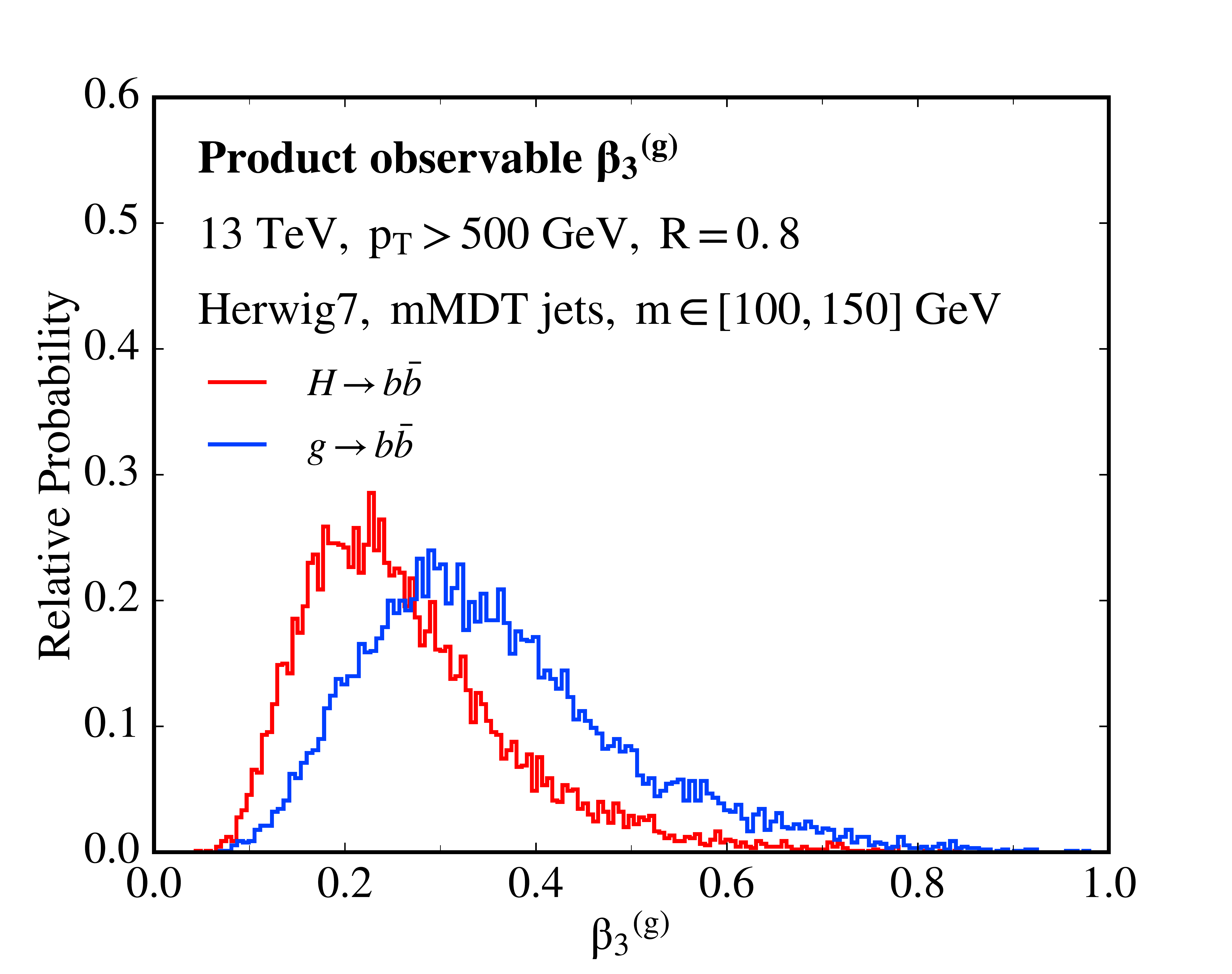}
		}\\
	\end{center}
	\caption{
		Distributions of the product observable for signal (red) and background (blue), measured on the samples of ungroomed (a) and groomed (b) jets showered with Herwig.  
	}
	\label{fig:herwig_prodobs}
\end{figure}

In \Fig{fig:prodobs}, we plot the distribution of these new product observables measured on signal and background jets showered in Pythia.  This shows that these product observables on ungroomed and groomed jets effectively separate signal from background.  Additionally, in \Fig{fig:herwig_prodobs}, we measure these product observables determined from the Pythia signal and background samples on the jets showered with Herwig. We observe a similar relative separation between the distributions, although the absolute scale is different, in the Herwig samples suggesting that these observables are sensitive to real physics, and not idiosyncrasies of the parton shower programs.

\begin{figure}[]
	\begin{center}
		\includegraphics[width=.3\textwidth]{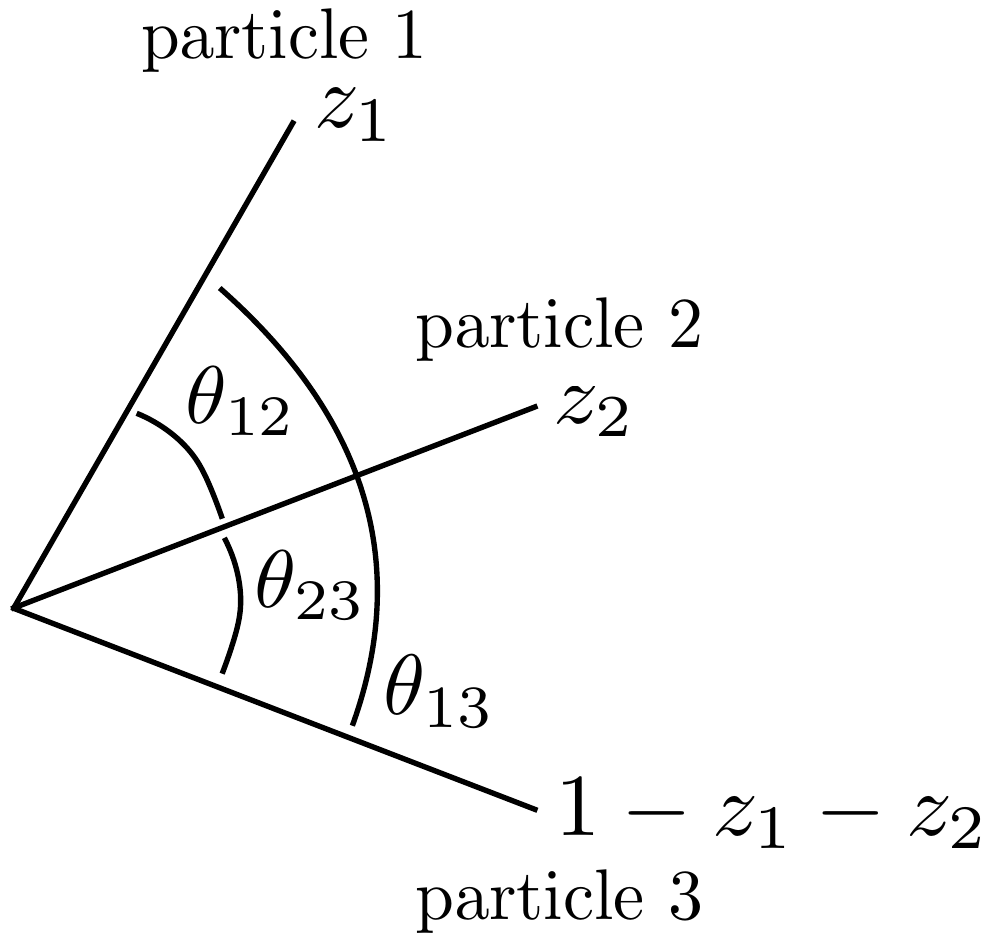}
		
		\caption{
			Illustration of the momentum fraction and pairwise angle variables that describe 3-body phase space.
		}
		\label{fig:spaces}
	\end{center}
\end{figure}

Especially for groomed jets, these simple forms for the product observables enable a nice interpretation of the physics to which they are sensitive.  In the case of ungroomed jets, there are multiple sources of radiation (final state, initial state, underlying event, etc.) that makes an interpretation a bit more challenging, so we won't discuss it more here.  When the jets are groomed with mMDT, however, contamination radiation from the initial state or underlying event is dominantly removed, and so a picture of the jet exclusively with radiation from the final state is accurate.  In this case, the mMDT jet with resolved 3-body phase space consists of the $b$ and $\bar b$ pair, and the dominant gluon emitted off of them.  The 3-body phase space configuration is shown in \Fig{fig:spaces}, with transverse momentum fractions $z_i$ and pairwise angles $\theta_{ij}$.  In what follows, we will let particles 1 and 2 be the $b$ and $\bar b$, and particle 3 be the gluon.

Because we make a cut on the jet mass and there is no soft singularity for $g\to b\bar b$ splitting, we assume that the emitted gluon is relatively soft and/or collinear with respect to the $b$ and the $\bar b$.  With this assumption, then the value of $\tau_1^{(0.5)}$ is completely determined by the $b$ and $\bar b$.  Then, the value of $\tau_1^{(0.5)}$ is approximately
\begin{align}
\tau_1^{(0.5)} &\simeq z((1-z)\theta_{12})^{0.5} +  (1-z)(z\theta_{12})^{0.5}\\
&=(z (1-z) \theta_{12}^2)^{0.25}\left(z^{0.75}(1-z)^{0.25}+z^{0.25}(1-z)^{0.75}\right) \,. \nonumber
\end{align}
Here, $z$ is the transverse momentum fraction of the $b$ quark subjet, for example.  The combination $z (1-z) \theta_{12}^2$ is just the ratio of the jet mass to the jet transverse momentum to this order, $m_J^2/p_{TJ}^2$, and is approximately constant because the mass cut is relatively narrow.  The term in parentheses on the right, $\left(z^{0.75}(1-z)^{0.25}+z^{0.25}(1-z)^{0.75}\right)$, is typically an order-1 number, as there is no soft singularity for $g\to b\bar b$ splitting nor for $H\to b\bar b$ decays.  So, to good approximation, $\tau_1^{(0.5)}$ on these jets with a mass cut is just some constant value.

The remaining factor in $\beta_3^{(g)}$ however, contains significantly interesting physics.  The two other $N$-subjettinesses that appear in that observable can be expressed as (see Ref.~\cite{Datta:2017rhs} for more details):
\begin{align}
&\tau_2^{(1)}= \frac{2z_3z_i}{z_i+z_3} \theta_{i3}\,,
&\tau_2^{(2)}= \frac{z_3z_i}{z_i+z_3} \theta_{i3}^2\,.
\end{align}
In writing this, we assume that the gluon, with transverse momentum fraction $z_3=1-z_1-z_2$, is the first particle clustered by the $k_T$ algorithm, and therefore sets the value of these $\tau_2$ observables.  $z_i$ is the transverse momentum fraction of the closer in angle of particles 1 or 2 (the $b$ or $\bar b$), with $\theta_{i3}$ this angle.  The ratio that appears in $\beta_3^{(g)}$ is therefore
\begin{equation}
\frac{\tau_2^{(2)}}{\tau_2^{(1)}} = \frac{\min[\theta_{13},\theta_{23}]}{2}\,.
\end{equation}
Therefore, with these assumptions, the groomed jet product observable is approximately proportional to the angle between the dominant gluon emission and the closer of the $b$ or $\bar b$:
\begin{equation}
\beta_3^{(g)} \propto \min[\theta_{13},\theta_{23}]\,.
\end{equation}
As color octets, gluons preferably emit at wide angles, while singlet Higgs bosons emit at small angles, and so we do expect this observable to provide discrimination power.

\subsection{Comparison to Standard Observables}\label{subsec:obscomp}
\begin{figure}[]
	\begin{center}
		\subfloat[]{\label{fig:nSD_comp_pythia}
			\includegraphics[width=7.2cm]{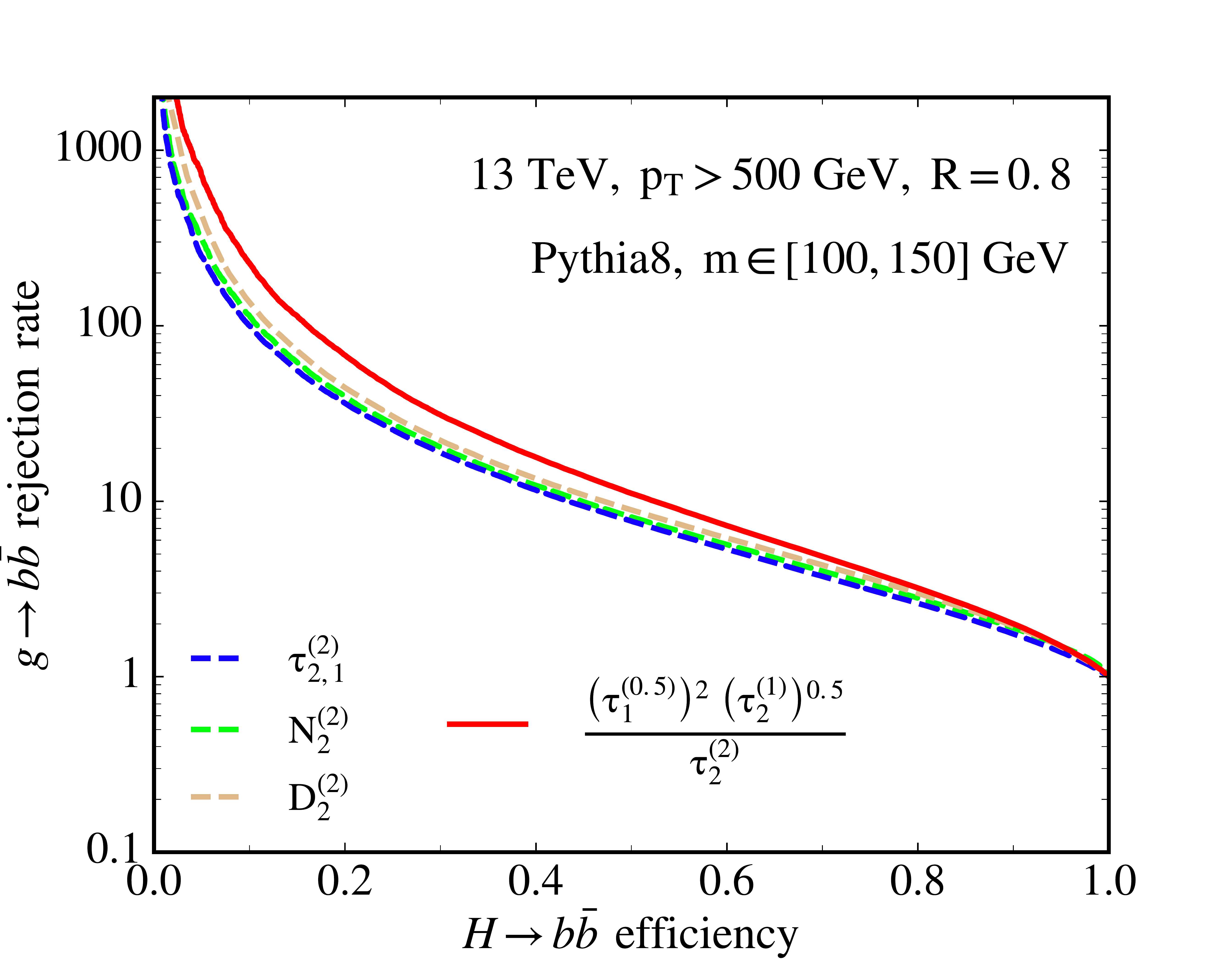}    
		}\qquad
		\subfloat[]{\label{fig:SD_comp_pythia}
			\includegraphics[width=7.2cm]{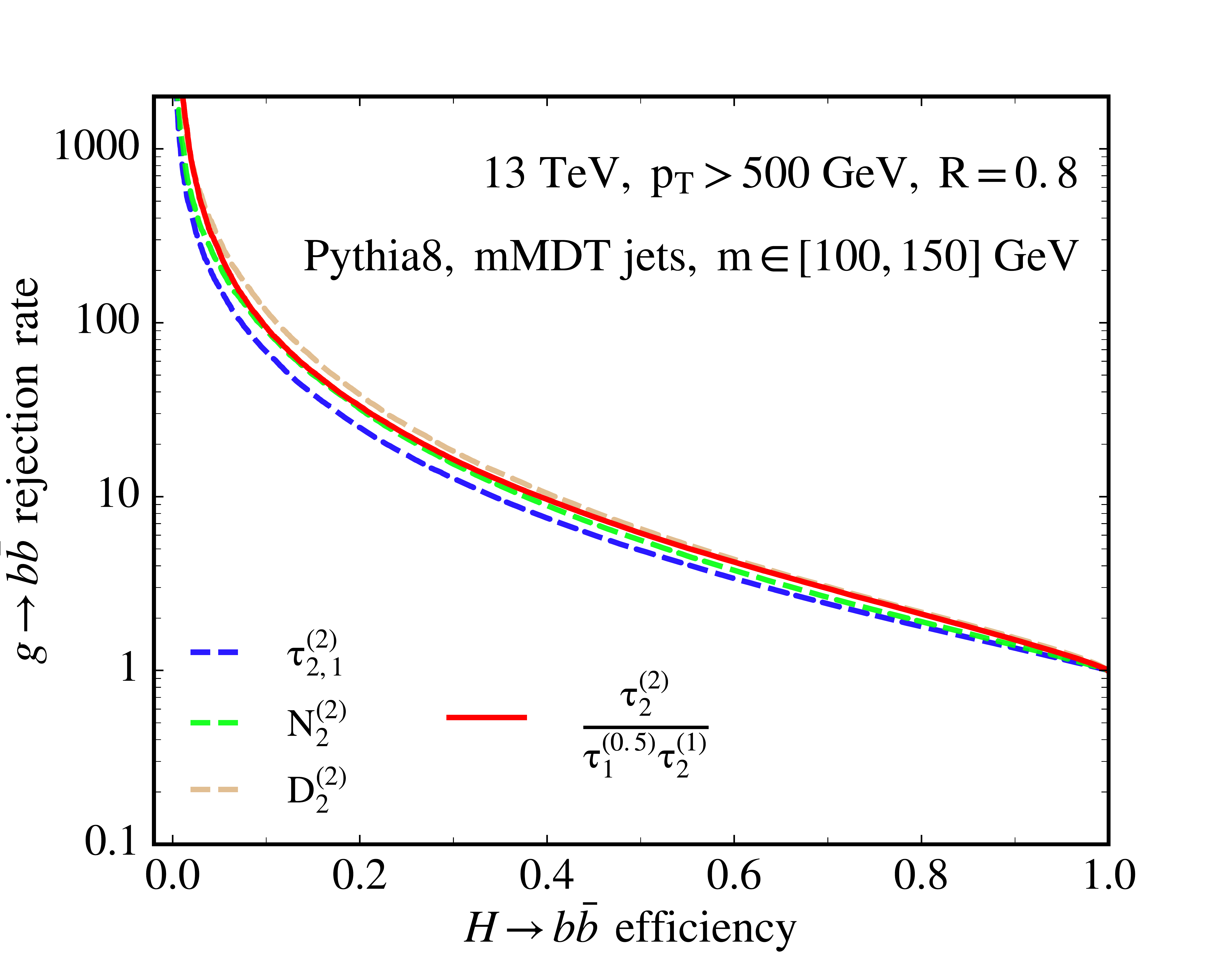}
		}\\
	\end{center}
	\caption{
		Signal efficiency versus background rejection rate for $N$-subjettiness ratio $\tau_{2,1}^{(2)}$, $N_{2}^{(2)}$, and $D_{2}^{(2)}$, measured on ungroomed (a) and groomed (b) jets showered using Pythia, compared to the discrimination power of the product observable $\beta_3$ or $\beta_3^{(g)}$. The discrimination power of the product observable is comparable to that of the standard observables.
	}\label{fig:pythia_obscomp}
\end{figure}

\begin{figure}[]
	\begin{center}
		
		\subfloat[]{\label{fig:nSD_comp_herwig}
			\includegraphics[width=7.2cm]{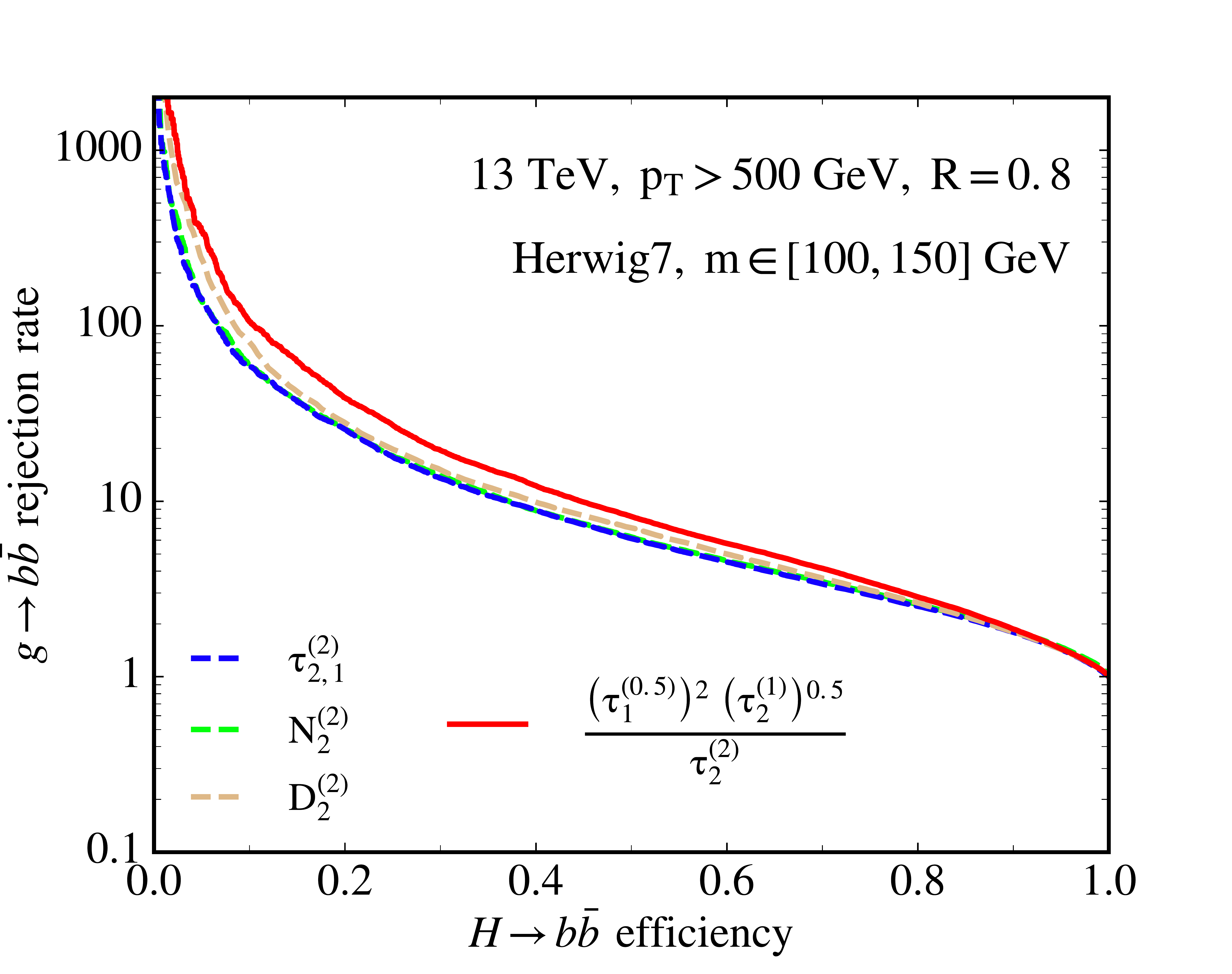}    
		}\qquad
		\subfloat[]{\label{fig:SD_comp_herwig}
			\includegraphics[width=7.2cm]{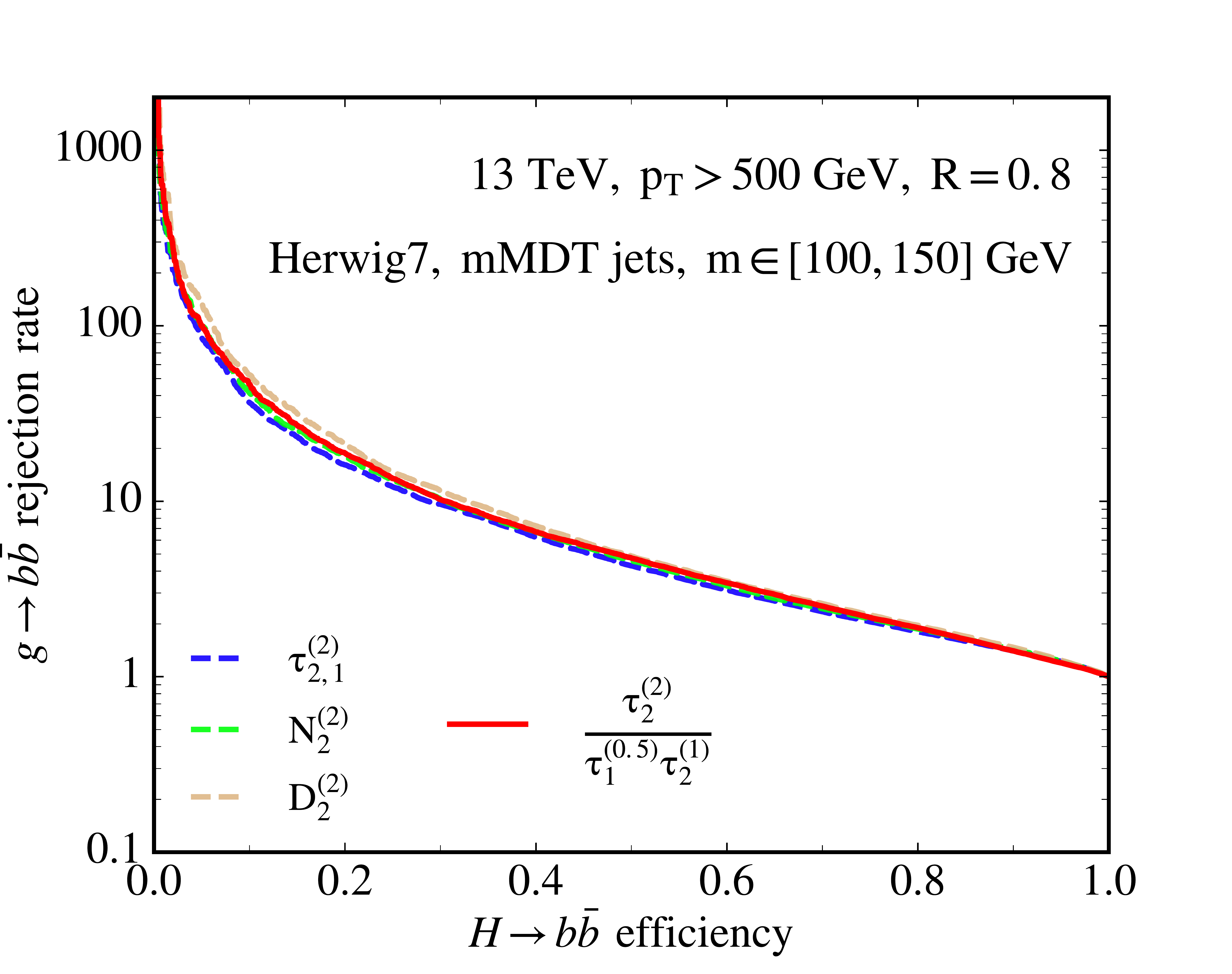}
		}\\
	\end{center}
	\caption{
		Signal efficiency versus background rejection rate for $N$-subjettiness ratio $\tau_{2,1}^{(2)}$, $N_{2}^{(2)}$, and $D_{2}^{(2)}$, measured on ungroomed (a) and groomed (b) jets showered using Herwig, compared to the discrimination power of the product observable $\beta_3$ or $\beta_3^{(g)}$.
	}
	\label{fig:herwig_obscomp}
\end{figure}

To demonstrate that these observables learned by the machine are indeed powerful, we compare their discrimination power to that of a collection of standard observables.  For comparison, we use $N$-subjettiness ratio $\tau_{2,1}^{(2)}$ with winner-take-all axes \cite{Bertolini:2013iqa,Larkoski:2014uqa,Larkoski:2014bia} and (generalized) energy correlation function ratios $D_2^{(2)}$ \cite{Larkoski:2014gra} and $N_2^{(2)}$ \cite{Moult:2016cvt}.  While these and related observables have been used for identification of boosted $H\to b\bar b$ decays, they are not necessarily optimized for this purpose.  Nevertheless, they provide a useful benchmark.  The signal efficiency versus background rejection rates for jets showered in Pythia are shown in \Fig{fig:pythia_obscomp} and for jets from Herwig, in \Fig{fig:herwig_obscomp}.  Most interestingly, for ungroomed jets, the new product observable $\beta_3$ outperforms each of these standard observables.  On groomed jets, the discrimination power of all of the observables is much closer and $D_2^{(2)}$ apparently slightly outperforms $\beta_3^{(g)}$.  Nevertheless, this demonstrates that, with very little human input, powerful discrimination observables can be constructed from what the machine learns.

\begin{figure}[!t]
	\begin{center}
		\subfloat[]{\label{fig:scatug}
			\includegraphics[width=7.6cm]{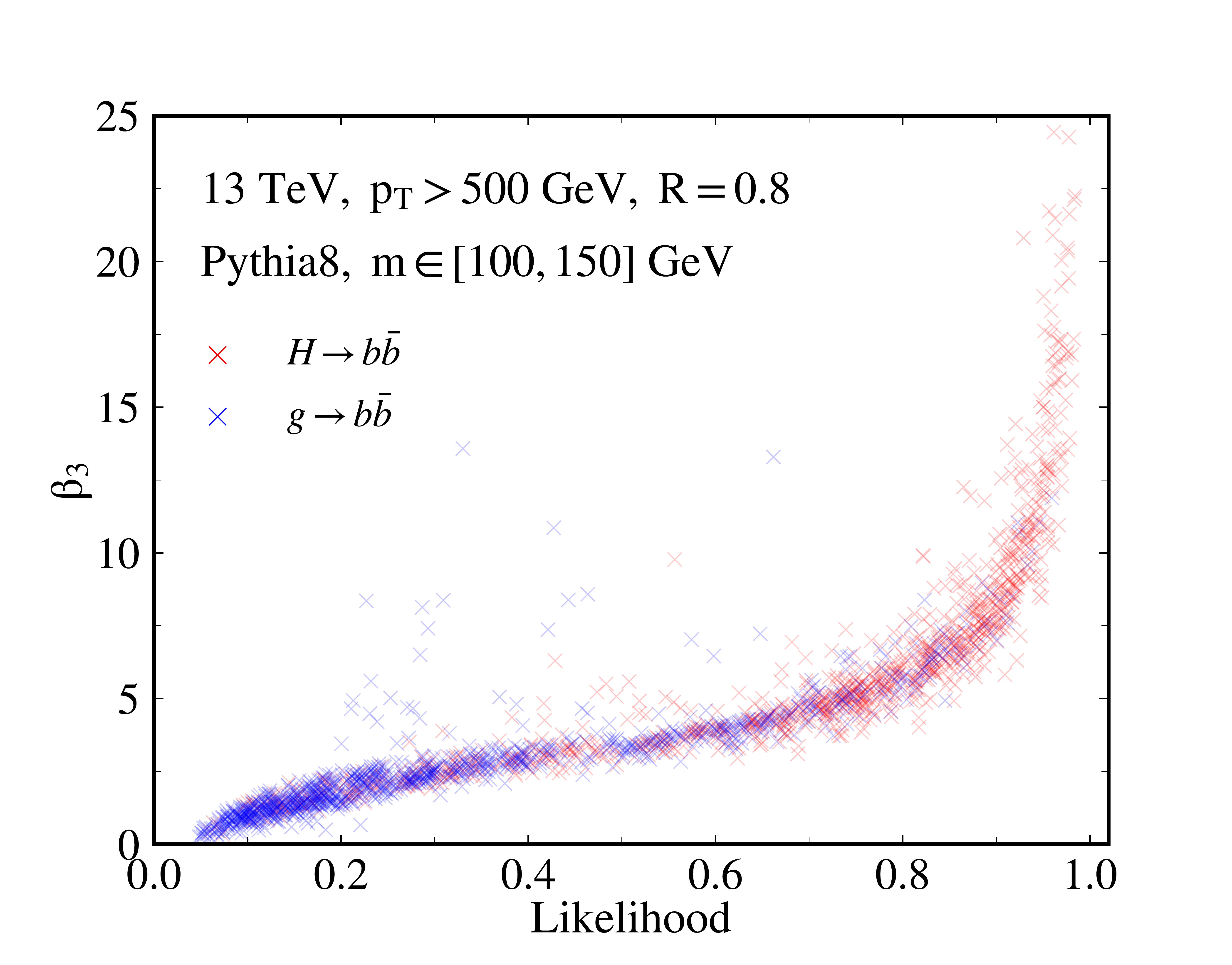}    
		}
		\subfloat[]{\label{fig:scatsd}
			\includegraphics[width=7.6cm]{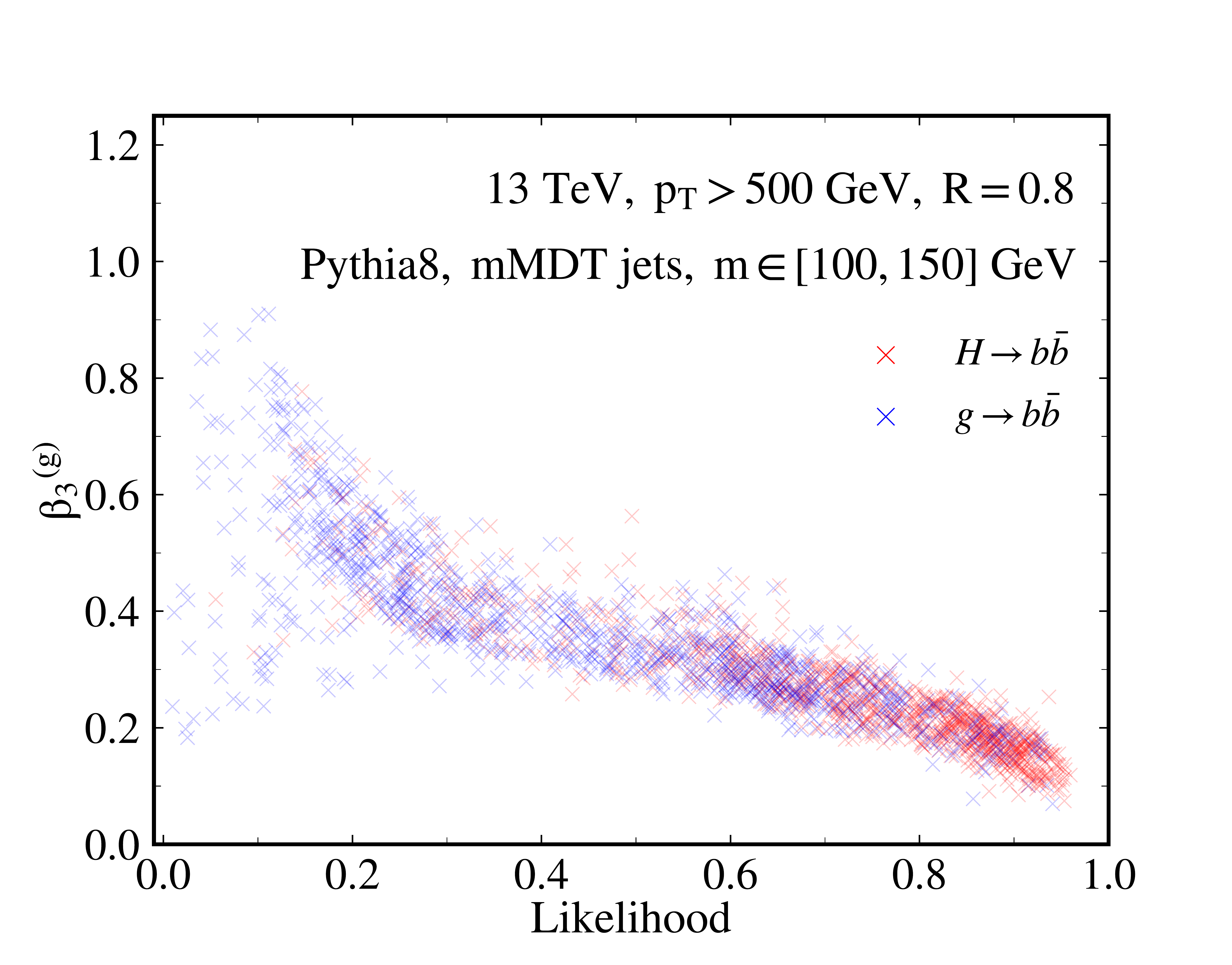}
		}
	\end{center}
	\caption{
		Scatter plots of 1000 signal (red) and background (blue) jets in the plane of $\beta_3$ versus the 3-body phase space likelihood ratio from the neural network.   Ungroomed jets are shown in (a), and mMDT groomed jets in (b).  The gross monotonic relationship between $\beta_3$ and the likelihood indicates that most of the information in the likelihood is captured in $\beta_3$, while the spread indicates that there is some discrimination information that is missed in $\beta_3$.
	}
	\label{fig:scatplots}
\end{figure}

It is instructive to also directly compare the value of our $\beta_3$ observables directly to the output likelihood ratio as determined on the 3-body phase space observables from the neural network.  In \Fig{fig:scatplots}, we have made scatter plots of the value of $\beta_3$ or $\beta_3^{(g)}$ (as appropriate) versus the likelihood ratio as measured on 1000 of both signal and background jets.  If the $\beta_3$ observables perfectly captured the information in the likelihood, these scatter plots should reduce to a monotonic curve.  The deviation from monotonicity is a measure of the information that $\beta_3$ misses with respect to the likelihood.  Broadly, these plots demonstrate a monotonic relationship between $\beta_3$ and the likelihood, but there is some spread.  The relative size of the spread is less for $\beta_3$ measured on ungroomed jets, which may reflect that in this case, $\beta_3$ captures more of the information in the likelihood than $\beta_3^{(g)}$.

\section{Conclusions}\label{sec:conc}

Previous deep learning studies in jet physics have shown immense promise. While it has been shown that appropriately designed deep learning techniques can outperform standard observables, such studies have not effectively probed what more information the machines are identifying.  Building on \Ref{Datta:2017rhs}, we propose a procedure that develops powerful new observables from the knowledge of the information contained in jets that contributes to an observable's discrimination power.
By systematically controlling the information fed to neural networks, it is possible to identify the minimal amount of information required to effectively discriminate between highly boosted decays of different massive particles and light QCD partons.  The method of planing introduced in Ref.~\cite{Chang:2017kvc} may also be useful in identifying the minimal information necessary for powerful discrimination.

Here, we have proposed an algorithm that can, in principle, be automated to construct new observables for any discrimination problem.  While the $H\to b\bar b$ application shows promise, such a procedure might also be applied to other specific problems like identifying $q$ vs.~$g$ jets, top quarks, or even to develop new observables that work effectively as generic anti-QCD taggers. The most improvement to this method  would be accomplished by construction of an optimal basis of functions with parameters that can be tuned to maximize discrimination power.  Of course, at this point, one is faced with a trade-off between simplicity and efficacy that must be taken into account.  By utilizing constructive deep learning techniques that are sensitive to exotic configurations within jets, this approach is presented with an intention to open the door to a whole new class of powerful substructure observables that can be tailored to specific or generic classification problems, while also providing further physics insight regarding the jets being studied.

\acknowledgments

We thank Gregor Kasieczka, Ian Moult, and Ben Nachman for comments on the manuscript.  The work of K.~D.~was supported by the Reed College Physics Department's Delord-Mockett Fund and Reed College Summer Scholarship Funds.

\bibliography{H2bb}

\providecommand{\href}[2]{#2}\begingroup\raggedright\begin{thebibliography}{10}

\bibitem{Cogan:2014oua}
J.~Cogan, M.~Kagan, E.~Strauss, and A.~Schwarztman, {\it {Jet-Images: Computer
  Vision Inspired Techniques for Jet Tagging}},  {\em JHEP} {\bf 02} (2015)
  118, [\href{http://arxiv.org/abs/1407.5675}{{\tt arXiv:1407.5675}}].

\bibitem{Almeida:2015jua}
L.~G. Almeida, M.~Backovic, M.~Cliche, S.~J. Lee, and M.~Perelstein, {\it
  {Playing Tag with ANN: Boosted Top Identification with Pattern Recognition}},
   {\em JHEP} {\bf 07} (2015) 086, [\href{http://arxiv.org/abs/1501.05968}{{\tt
  arXiv:1501.05968}}].

\bibitem{deOliveira:2015xxd}
L.~de~Oliveira, M.~Kagan, L.~Mackey, B.~Nachman, and A.~Schwartzman, {\it
  {Jet-images ? deep learning edition}},  {\em JHEP} {\bf 07} (2016) 069,
  [\href{http://arxiv.org/abs/1511.05190}{{\tt arXiv:1511.05190}}].

\bibitem{Baldi:2016fql}
P.~Baldi, K.~Bauer, C.~Eng, P.~Sadowski, and D.~Whiteson, {\it {Jet
  Substructure Classification in High-Energy Physics with Deep Neural
  Networks}},  {\em Phys. Rev.} {\bf D93} (2016), no.~9 094034,
  [\href{http://arxiv.org/abs/1603.09349}{{\tt arXiv:1603.09349}}].

\bibitem{Guest:2016iqz}
D.~Guest, J.~Collado, P.~Baldi, S.-C. Hsu, G.~Urban, and D.~Whiteson, {\it {Jet
  Flavor Classification in High-Energy Physics with Deep Neural Networks}},
  {\em Phys. Rev.} {\bf D94} (2016), no.~11 112002,
  [\href{http://arxiv.org/abs/1607.08633}{{\tt arXiv:1607.08633}}].

\bibitem{Conway:2016caq}
J.~S. Conway, R.~Bhaskar, R.~D. Erbacher, and J.~Pilot, {\it {Identification of
  High-Momentum Top Quarks, Higgs Bosons, and W and Z Bosons Using Boosted
  Event Shapes}},  {\em Phys. Rev.} {\bf D94} (2016), no.~9 094027,
  [\href{http://arxiv.org/abs/1606.06859}{{\tt arXiv:1606.06859}}].

\bibitem{Barnard:2016qma}
J.~Barnard, E.~N. Dawe, M.~J. Dolan, and N.~Rajcic, {\it {Parton Shower
  Uncertainties in Jet Substructure Analyses with Deep Neural Networks}},  {\em
  Phys. Rev.} {\bf D95} (2017), no.~1 014018,
  [\href{http://arxiv.org/abs/1609.00607}{{\tt arXiv:1609.00607}}].

\bibitem{Komiske:2016rsd}
P.~T. Komiske, E.~M. Metodiev, and M.~D. Schwartz, {\it {Deep learning in
  color: towards automated quark/gluon jet discrimination}},  {\em JHEP} {\bf
  01} (2017) 110, [\href{http://arxiv.org/abs/1612.01551}{{\tt
  arXiv:1612.01551}}].

\bibitem{deOliveira:2017pjk}
L.~de~Oliveira, M.~Paganini, and B.~Nachman, {\it {Learning Particle Physics by
  Example: Location-Aware Generative Adversarial Networks for Physics
  Synthesis}},  \href{http://arxiv.org/abs/1701.05927}{{\tt arXiv:1701.05927}}.

\bibitem{Kasieczka:2017nvn}
G.~Kasieczka, T.~Plehn, M.~Russell, and T.~Schell, {\it {Deep-learning Top
  Taggers or The End of QCD?}},  {\em JHEP} {\bf 05} (2017) 006,
  [\href{http://arxiv.org/abs/1701.08784}{{\tt arXiv:1701.08784}}].

\bibitem{Louppe:2017ipp}
G.~Louppe, K.~Cho, C.~Becot, and K.~Cranmer, {\it {QCD-Aware Recursive Neural
  Networks for Jet Physics}},  \href{http://arxiv.org/abs/1702.00748}{{\tt
  arXiv:1702.00748}}.

\bibitem{Dery:2017fap}
L.~M. Dery, B.~Nachman, F.~Rubbo, and A.~Schwartzman, {\it {Weakly Supervised
  Classification in High Energy Physics}},
  \href{http://arxiv.org/abs/1702.00414}{{\tt arXiv:1702.00414}}.

\bibitem{Pearkes:2017hku}
J.~Pearkes, W.~Fedorko, A.~Lister, and C.~Gay, {\it {Jet Constituents for Deep
  Neural Network Based Top Quark Tagging}},
  \href{http://arxiv.org/abs/1704.02124}{{\tt arXiv:1704.02124}}.

\bibitem{Cohen:2017exh}
T.~Cohen, M.~Freytsis, and B.~Ostdiek, {\it {(Machine) Learning to Do More with
  Less}},  \href{http://arxiv.org/abs/1706.09451}{{\tt arXiv:1706.09451}}.

\bibitem{Butter:2017cot}
A.~Butter, G.~Kasieczka, T.~Plehn, and M.~Russell, {\it {Deep-learned Top
  Tagging using Lorentz Invariance and Nothing Else}},
  \href{http://arxiv.org/abs/1707.08966}{{\tt arXiv:1707.08966}}.

\bibitem{Metodiev:2017vrx}
E.~M. Metodiev, B.~Nachman, and J.~Thaler, {\it {Classification without labels:
  Learning from mixed samples in high energy physics}},
  \href{http://arxiv.org/abs/1708.02949}{{\tt arXiv:1708.02949}}.

\bibitem{Chang:2017kvc}
S.~Chang, T.~Cohen, and B.~Ostdiek, {\it {What is the Machine Learning?}},
  \href{http://arxiv.org/abs/1709.10106}{{\tt arXiv:1709.10106}}.

\bibitem{Larkoski:2017jix}
A.~J. Larkoski, I.~Moult, and B.~Nachman, {\it {Jet Substructure at the Large
  Hadron Collider: A Review of Recent Advances in Theory and Machine
  Learning}},  \href{http://arxiv.org/abs/1709.04464}{{\tt arXiv:1709.04464}}.

\bibitem{Datta:2017rhs}
K.~Datta and A.~Larkoski, {\it {How Much Information is in a Jet?}},  {\em
  JHEP} {\bf 06} (2017) 073, [\href{http://arxiv.org/abs/1704.08249}{{\tt
  arXiv:1704.08249}}].

\bibitem{Aguilar-Saavedra:2017rzt}
J.~A. Aguilar-Saavedra, J.~H. Collins, and R.~K. Mishra, {\it {A generic
  anti-QCD jet tagger}},  \href{http://arxiv.org/abs/1709.01087}{{\tt
  arXiv:1709.01087}}.

\bibitem{CMS:2017cbv}
{\bf CMS} Collaboration, C.~Collaboration, {\it {Inclusive search for the
  standard model Higgs boson produced in pp collisions at
  $\sqrt{s}=13~\mathrm{TeV}$ using H$\rightarrow \mathrm{b\bar{\mathrm{b}}}$
  decays}}, .

\bibitem{Thaler:2010tr}
J.~Thaler and K.~Van~Tilburg, {\it {Identifying Boosted Objects with
  N-subjettiness}},  {\em JHEP} {\bf 03} (2011) 015,
  [\href{http://arxiv.org/abs/1011.2268}{{\tt arXiv:1011.2268}}].

\bibitem{Thaler:2011gf}
J.~Thaler and K.~Van~Tilburg, {\it {Maximizing Boosted Top Identification by
  Minimizing N-subjettiness}},  {\em JHEP} {\bf 02} (2012) 093,
  [\href{http://arxiv.org/abs/1108.2701}{{\tt arXiv:1108.2701}}].

\bibitem{Dasgupta:2013ihk}
M.~Dasgupta, A.~Fregoso, S.~Marzani, and G.~P. Salam, {\it {Towards an
  understanding of jet substructure}},  {\em JHEP} {\bf 09} (2013) 029,
  [\href{http://arxiv.org/abs/1307.0007}{{\tt arXiv:1307.0007}}].

\bibitem{Dasgupta:2013via}
M.~Dasgupta, A.~Fregoso, S.~Marzani, and A.~Powling, {\it {Jet substructure
  with analytical methods}},  {\em Eur. Phys. J.} {\bf C73} (2013), no.~11
  2623, [\href{http://arxiv.org/abs/1307.0013}{{\tt arXiv:1307.0013}}].

\bibitem{Stewart:2010tn}
I.~W. Stewart, F.~J. Tackmann, and W.~J. Waalewijn, {\it {N-Jettiness: An
  Inclusive Event Shape to Veto Jets}},  {\em Phys. Rev. Lett.} {\bf 105}
  (2010) 092002, [\href{http://arxiv.org/abs/1004.2489}{{\tt
  arXiv:1004.2489}}].

\bibitem{Larkoski:2013eya}
A.~J. Larkoski, G.~P. Salam, and J.~Thaler, {\it {Energy Correlation Functions
  for Jet Substructure}},  {\em JHEP} {\bf 06} (2013) 108,
  [\href{http://arxiv.org/abs/1305.0007}{{\tt arXiv:1305.0007}}].

\bibitem{Catani:1993hr}
S.~Catani, Y.~L. Dokshitzer, M.~H. Seymour, and B.~R. Webber, {\it
  {Longitudinally invariant $K_t$ clustering algorithms for hadron hadron
  collisions}},  {\em Nucl. Phys.} {\bf B406} (1993) 187--224.

\bibitem{Ellis:1993tq}
S.~D. Ellis and D.~E. Soper, {\it {Successive combination jet algorithm for
  hadron collisions}},  {\em Phys. Rev.} {\bf D48} (1993) 3160--3166,
  [\href{http://arxiv.org/abs/hep-ph/9305266}{{\tt hep-ph/9305266}}].

\bibitem{Blazey:2000qt}
G.~C. Blazey et~al., {\it {Run II jet physics}},  in {\em {QCD and weak boson
  physics in Run II. Proceedings, Batavia, USA, March 4-6, June 3-4, November
  4-6, 1999}}, pp.~47--77, 2000.
\newblock \href{http://arxiv.org/abs/hep-ex/0005012}{{\tt hep-ex/0005012}}.

\bibitem{Dokshitzer:1997in}
Y.~L. Dokshitzer, G.~D. Leder, S.~Moretti, and B.~R. Webber, {\it {Better jet
  clustering algorithms}},  {\em JHEP} {\bf 08} (1997) 001,
  [\href{http://arxiv.org/abs/hep-ph/9707323}{{\tt hep-ph/9707323}}].

\bibitem{Wobisch:1998wt}
M.~Wobisch and T.~Wengler, {\it {Hadronization corrections to jet
  cross-sections in deep inelastic scattering}},  in {\em {Monte Carlo
  generators for HERA physics. Proceedings, Workshop, Hamburg, Germany,
  1998-1999}}, pp.~270--279, 1998.
\newblock \href{http://arxiv.org/abs/hep-ph/9907280}{{\tt hep-ph/9907280}}.

\bibitem{Adloff:2000tq}
{\bf H1} Collaboration, C.~Adloff et~al., {\it {Measurement and QCD analysis of
  jet cross-sections in deep inelastic positron - proton collisions at s**(1/2)
  of 300-GeV}},  {\em Eur. Phys. J.} {\bf C19} (2001) 289--311,
  [\href{http://arxiv.org/abs/hep-ex/0010054}{{\tt hep-ex/0010054}}].

\bibitem{Alwall:2014hca}
J.~Alwall, R.~Frederix, S.~Frixione, V.~Hirschi, F.~Maltoni, O.~Mattelaer,
  H.~S. Shao, T.~Stelzer, P.~Torrielli, and M.~Zaro, {\it {The automated
  computation of tree-level and next-to-leading order differential cross
  sections, and their matching to parton shower simulations}},  {\em JHEP} {\bf
  07} (2014) 079, [\href{http://arxiv.org/abs/1405.0301}{{\tt
  arXiv:1405.0301}}].

\bibitem{Sjostrand:2006za}
T.~Sjostrand, S.~Mrenna, and P.~Z. Skands, {\it {PYTHIA 6.4 Physics and
  Manual}},  {\em JHEP} {\bf 05} (2006) 026,
  [\href{http://arxiv.org/abs/hep-ph/0603175}{{\tt hep-ph/0603175}}].

\bibitem{Sjostrand:2014zea}
T.~Sj�strand, S.~Ask, J.~R. Christiansen, R.~Corke, N.~Desai, P.~Ilten,
  S.~Mrenna, S.~Prestel, C.~O. Rasmussen, and P.~Z. Skands, {\it {An
  Introduction to PYTHIA 8.2}},  {\em Comput. Phys. Commun.} {\bf 191} (2015)
  159--177, [\href{http://arxiv.org/abs/1410.3012}{{\tt arXiv:1410.3012}}].

\bibitem{Bahr:2008pv}
M.~Bahr et~al., {\it {Herwig++ Physics and Manual}},  {\em Eur. Phys. J.} {\bf
  C58} (2008) 639--707, [\href{http://arxiv.org/abs/0803.0883}{{\tt
  arXiv:0803.0883}}].

\bibitem{Bellm:2015jjp}
J.~Bellm et~al., {\it {Herwig 7.0/Herwig++ 3.0 release note}},  {\em Eur. Phys.
  J.} {\bf C76} (2016), no.~4 196, [\href{http://arxiv.org/abs/1512.01178}{{\tt
  arXiv:1512.01178}}].

\bibitem{Cacciari:2011ma}
M.~Cacciari, G.~P. Salam, and G.~Soyez, {\it {FastJet User Manual}},  {\em Eur.
  Phys. J.} {\bf C72} (2012) 1896, [\href{http://arxiv.org/abs/1111.6097}{{\tt
  arXiv:1111.6097}}].

\bibitem{Cacciari:2005hq}
M.~Cacciari and G.~P. Salam, {\it {Dispelling the $N^{3}$ myth for the $k_t$
  jet-finder}},  {\em Phys. Lett.} {\bf B641} (2006) 57--61,
  [\href{http://arxiv.org/abs/hep-ph/0512210}{{\tt hep-ph/0512210}}].

\bibitem{Cacciari:2008gp}
M.~Cacciari, G.~P. Salam, and G.~Soyez, {\it {The Anti-k(t) jet clustering
  algorithm}},  {\em JHEP} {\bf 04} (2008) 063,
  [\href{http://arxiv.org/abs/0802.1189}{{\tt arXiv:0802.1189}}].

\bibitem{chollet2015keras}
F.~Chollet, ``Keras.'' \url{https://github.com/fchollet/keras}, 2015.

\bibitem{Neyman289}
J.~Neyman and E.~S. Pearson, {\it On the problem of the most efficient tests of
  statistical hypotheses},  {\em Philosophical Transactions of the Royal
  Society of London A: Mathematical, Physical and Engineering Sciences} {\bf
  231} (1933), no.~694-706 289--337,
  [\href{http://arxiv.org/abs/http://rsta.royalsocietypublishing.org/content/231/694-706/289.full.pdf}{{\tt
  http://rsta.royalsocietypublishing.org/content/231/694-706/289.full.pdf}}].

\bibitem{Larkoski:2013paa}
A.~J. Larkoski and J.~Thaler, {\it {Unsafe but Calculable: Ratios of
  Angularities in Perturbative QCD}},  {\em JHEP} {\bf 09} (2013) 137,
  [\href{http://arxiv.org/abs/1307.1699}{{\tt arXiv:1307.1699}}].

\bibitem{Larkoski:2015lea}
A.~J. Larkoski, S.~Marzani, and J.~Thaler, {\it {Sudakov Safety in Perturbative
  QCD}},  {\em Phys. Rev.} {\bf D91} (2015), no.~11 111501,
  [\href{http://arxiv.org/abs/1502.01719}{{\tt arXiv:1502.01719}}].

\bibitem{Bertolini:2013iqa}
D.~Bertolini, T.~Chan, and J.~Thaler, {\it {Jet Observables Without Jet
  Algorithms}},  {\em JHEP} {\bf 04} (2014) 013,
  [\href{http://arxiv.org/abs/1310.7584}{{\tt arXiv:1310.7584}}].

\bibitem{Larkoski:2014uqa}
A.~J. Larkoski, D.~Neill, and J.~Thaler, {\it {Jet Shapes with the Broadening
  Axis}},  {\em JHEP} {\bf 04} (2014) 017,
  [\href{http://arxiv.org/abs/1401.2158}{{\tt arXiv:1401.2158}}].

\bibitem{Larkoski:2014bia}
A.~J. Larkoski and J.~Thaler, {\it {Aspects of jets at 100 TeV}},  {\em Phys.
  Rev.} {\bf D90} (2014), no.~3 034010,
  [\href{http://arxiv.org/abs/1406.7011}{{\tt arXiv:1406.7011}}].

\bibitem{Larkoski:2014gra}
A.~J. Larkoski, I.~Moult, and D.~Neill, {\it {Power Counting to Better Jet
  Observables}},  {\em JHEP} {\bf 12} (2014) 009,
  [\href{http://arxiv.org/abs/1409.6298}{{\tt arXiv:1409.6298}}].

\bibitem{Moult:2016cvt}
I.~Moult, L.~Necib, and J.~Thaler, {\it {New Angles on Energy Correlation
  Functions}},  {\em JHEP} {\bf 12} (2016) 153,
  [\href{http://arxiv.org/abs/1609.07483}{{\tt arXiv:1609.07483}}].

\end{thebibliography}\endgroup
\end{document}